\documentclass[showpacs,showkeys,byrevtex,twocolumn,pra]{revtex4}%
\usepackage{amsfonts}
\usepackage{amsmath}
\usepackage{amssymb}
\usepackage{graphicx}
\usepackage{verbatim}%
\providecommand{\U}[1]{\protect\rule{.1in}{.1in}}
\newtheorem{theorem}{Theorem}

\newenvironment{proof}[1][Proof]{\noindent\textbf{#1.} }{\ \rule{0.5em}{0.5em}}
\begin{document}
\preprint{ }
\title[ ]{Quantum Forbidden-Interval Theorems for Stochastic Resonance}
\author{Mark M. Wilde$^{1,2}$}
\email{mark.wilde@usc.edu}
\author{Bart Kosko$^{1}$}
\affiliation{$^{1}$Center for Quantum Information Science and Technology, Department of
Electrical Engineering, University of Southern California, Los Angeles,
California 90089, USA}
\affiliation{$^{2}$Hearne Institute for Theoretical Physics, Department of Physics and
Astronomy, Louisiana State University, Baton Rouge, Louisiana 70803}

\begin{abstract}
We extend the classical forbidden-interval theorems for a stochastic-resonance
noise benefit in a nonlinear system to a quantum-optical communication model
and a continuous-variable quantum key distribution model. Each quantum
forbidden-interval theorem gives a necessary and sufficient condition that
determines whether stochastic resonance occurs in quantum communication of
classical messages. The quantum theorems apply to any quantum noise source
that has finite variance or that comes from the family of infinite-variance
alpha-stable probability densities. Simulations show the noise
benefits for the basic quantum communication model and the continuous-variable
quantum key distribution model.

\end{abstract}
\volumeyear{2007}
\volumenumber{ }
\issuenumber{ }
\eid{ }
\date{\today}
\received{}

\revised{}

\accepted{}

\published{}

\startpage{1}
\endpage{ }
\maketitle

\begin{figure}
[ptb]
\begin{center}
\includegraphics[
height=2.1741in,
width=2.8911in
]%
{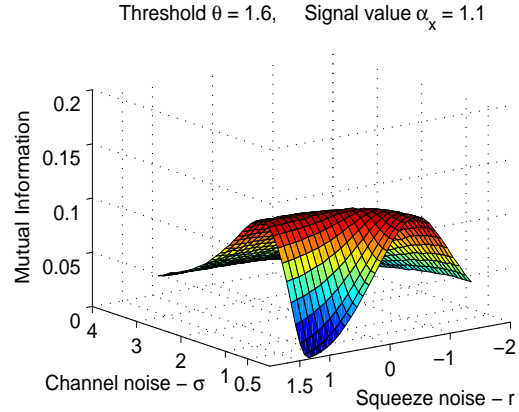}%
\caption{(Color online) Stochastic resonance in the basic quantum-optical
communication model with Gaussian noise. The sender Alice encodes coherent
states with amplitude $A=1.1$. The receiver Bob decodes with threshold
$\theta=1.6$. The graph shows the smoothed mutual information as a function of
the standard deviation $\sigma$ of the quantum Gaussian noise and the
squeezing strength $r$ for 100 simulation runs. Each run generated 10,000
input-output signal pairs to estimate the mutual information. The SR effect
occurs because the channel noise mean $\mu=0$ and thus $\mu$ lies outside the
forbidden interval $(.5,2.7)$.}%
\label{fig:basic-QSR-model}%
\end{center}
\end{figure}

Stochastic resonance (SR) occurs in a nonlinear system when noise benefits the
system
\cite{jpa1981benzi,nature1995wiesenfeld,phystoday1996bulsara,revmod1998gamma,noisebook}%
. SR can occur in both classical and quantum systems
\cite{revmod1998gamma,njp1999hanggi} that use noise to help detect faint
signals. The footprint of SR is a nonmonotonic curve that results when the
system performance measure depends on the intensity of the noise source.
Figure~\ref{fig:basic-QSR-model} shows such an SR surface for a
quantum-optical communication system with both additive channel noise and
squeezing noise. Mutual information measures the noise benefits in bits.

The classical SR forbidden-interval theorems give necessary and sufficient
conditions for an SR noise benefit in terms of mutual information
\cite{nn2003kosko,pre2004kosko} when the system nonlinear is a threshold. The
noise benefit turns on whether the noise mean or location $a$ lies in an
interval that depends on the threshold $\theta$ and the bipolar subthreshold
signals $A$ and $-A$: SR occurs if and only if $a\notin(\theta-A,\theta+A)$
where $-A<A<\theta$. This result holds for all finite-variance noise and all
infinite-variance stable noise. But it guarantees only that some SR noise
benefit occurs in the system for the given choice of parameters. SR stochastic
learning algorithms \cite{PhysRevE.64.051110} can then search for the optimal
noise level.

This paper generalizes the classical forbidden-interval theorems to a
quantum-optimal communication system that uses squeezed light
\cite{modopt1987knight,book2005gerry}. The corresponding quantum
forbidden-interval theorems give similar necessary and sufficient conditions
for a noise benefit but include the strength of light squeezing as a
parameter. The quantum-optical system in Figure~\ref{fig:basic-QSR-model}
produces SR because the noise mean is zero and so does not lie in the system's
forbidden interval $(.5,2.7)$. We also show that modified versions of the
quantum forbidden-interval theorems hold in continuous-variable quantum key
distribution with thresholding \cite{arxiv2000hirano,PhysRevA.67.022308}.%

\textit{Model for Quantum-Optical Thresholding System}---We first develop the
basic quantum-optical communication protocol. The first quantum
forbidden-interval theorem applies to this communication model. We present the
sender Alice's operations for encoding information, the effect of the noisy
quantum channel on the state that Alice transmits, and the receiver Bob's
detection scheme.

Alice wants to encode a message bit $S$ using quantum-optical techniques. The
protocol begins with Alice possessing a vacuum mode. We describe our model in
the Heisenberg picture. Let $\hat{x}$ denote the position-quadrature operator
of Alice's vacuum mode where $\hat{x}=\left(  \hat{a}+\hat{a}^{\dag}\right)
/\sqrt{2}$ and $\hat{a}$ is the annihilation operator for her vacuum mode
\cite{book2005gerry}. We consider only the position-quadrature operator's
evolution. Her vacuum state collapses to a zero-mean $1/2$-variance Gaussian
random variable $X$ if she measures it with an ideal position-quadrature
homodyne detector. Suppose that Alice does not measure it. Suppose instead
that she sends her mode through a position-quadrature squeezer. Suppose
further that she can control the strength of squeezing with a squeezing
parameter $r$. The position-quadrature squeezer is a unitary operator $\hat
{S}\left(  r\right)  \equiv\exp\left\{  r\left(  \hat{a}^{2}-\left(  \hat
{a}^{\dag}\right)  ^{2}\right)  \right\}  $ \cite{book2005gerry}. Her operator
$\hat{x}$ evolves under the squeezer as $\hat{S}^{\dag}\left(  r\right)
\hat{x}\hat{S}\left(  r\right)  =\hat{x}e^{-r}$. She encodes the random
message bit $S\in\left\{  0,1\right\}  $ by displacing her state by $\alpha
\in\mathbb{C}$ if $S=1$ or by $-\alpha$ if $S=0$. The displacement is a
unitary operator $\hat{D}\left(  \alpha\right)  \equiv\exp\left\{  \alpha
\hat{a}^{\dag}-\alpha^{\ast}\hat{a}\right\}  $ \cite{book2005gerry}. Let
$\alpha_{S}$ be the conditional displacement $\alpha_{S}=\left(  -1\right)
^{S+1}\alpha$. Her operator $\hat{x}e^{-r}$ evolves under the conditional
displacement $\hat{D}\left(  \alpha_{S}\right)  $ as%
\begin{equation}
\hat{D}^{\dag}\left(  \alpha_{S}\right)  \left(  \hat{x}e^{-r}\right)  \hat
{D}\left(  \alpha_{S}\right)  =\hat{x}e^{-r}+\left(  -1\right)  ^{S+1}%
\alpha_{x} \label{eq:qmessage}%
\end{equation}
where $\alpha_{x}=\operatorname{Re}\left\{  \alpha\right\}  $. This equality
gives the Heisenberg-picture observable that corresponds to Alice's mode
before she sends it over the noisy channel. The message bit $S$ appears as a
displacement in (\ref{eq:qmessage}).

Alice sends her mode to Bob over an additive noisy bosonic channel
\cite{pra2001holevo} that adds a random displacement $\nu\in\mathbb{C}$ to its
input state. The channel randomly displaces any annihilation operator $\hat
{a}$\ as $\hat{D}^{\dag}\left(  \nu\right)  \hat{a}\hat{D}\left(  \nu\right)
=\hat{a}+\nu$. This is the quantum-channel analogue to a classical continuous
additive noisy channel \cite{book1991cover}. The term $\hat{x}e^{-r}+\left(
-1\right)  ^{S+1}\alpha_{x}+\nu_{x}$\ is the Heisenberg-picture
position-quadrature observable that corresponds to the state that Bob receives
after Alice sends her mode over the noisy channel. Random variable $\nu
_{x}=\operatorname{Re}\left\{  \nu\right\}  $ and corresponds to the
position-quadrature noise.

Bob detects the information that Alice encodes by performing
position-quadrature homodyne detection with inefficient photodetectors. We
model this non-ideal homodyne detection as a lossy transmission through a
material with linear absorption \cite{PhysRevA.48.4598}. A beamsplitter with
transmittivity $\eta$ models the linear absorptive material. Then the
Heisenberg-picture observable after the lossy beamsplitter is%
\begin{equation}
\sqrt{\eta}\left(  \left(  -1\right)  ^{S+1}\alpha_{x}+\hat{x}e^{-r}+\nu
_{x}\right)  +\sqrt{1-\eta}\hat{x}_{H}%
\end{equation}
where $\eta$ is the quantum efficiency of the homodyne detection and $\hat
{x}_{H}$ is the position quadrature operator of an input vacuum mode. Bob
measures the position quadrature observable and the state collapses to the
random variable%
\begin{equation}
\sqrt{\eta}\left(  \left(  -1\right)  ^{S+1}\alpha_{x}+Xe^{-r}+\nu_{x}\right)
+\sqrt{1-\eta}X_{H}. \label{eq:bobs-state}%
\end{equation}
$X_{H}$ is a zero-mean $1/2$-variance Gaussian random variable that
corresponds to the vacuum observable $\hat{x}_{H}$. Random variables $Xe^{-r}%
$, $\nu_{x}$, and $X_{H}$\ are independent because random variable $Xe^{-r}$
comes from the vacuum fluctuations of Alice's original mode, because $\nu_{x}$
is Bob's loss of knowledge due to the state's propagation through a noisy
quantum channel, and because $X_{H}$ comes from the vacuum contributions of
non-ideal position-quadrature homodyne detection. Let random variable $N$ sum
all noise terms:%
\begin{equation}
N\equiv\sqrt{\eta}\left(  Xe^{-r}+\nu_{x}\right)  +\sqrt{1-\eta}X_{H}.
\label{eq:all-noise}%
\end{equation}
The density $p_{N}\left(  n\right)  $ of random variable $N$ is%
\begin{equation}
p_{N}\left(  n\right)  =\left(  p_{\sqrt{\eta}Xe^{-r}}\ast p_{\sqrt{\eta}%
\nu_{x}}\ast p_{\sqrt{1-\eta}X_{H}}\right)  \left(  n\right)
\label{eq:noise-density}%
\end{equation}
where $p_{\sqrt{\eta}Xe^{-r}}\left(  n\right)  $ is the density of a zero-mean
$\eta e^{-2r}/2$-variance Gaussian random variable, $p_{\sqrt{\eta}\nu_{x}%
}\left(  n\right)  $ is the density of $\sqrt{\eta}\nu_{x}$, $p_{\sqrt{1-\eta
}X_{H}}\left(  n\right)  $ is the density of a zero-mean $\left(
1-\eta\right)  /2$-variance Gaussian random variable, and $\ast$ denotes
convolution. The density $p_{N}\left(  n\right)  $ is a convolution because
random variables $Xe^{-r}$, $\nu_{x}$, and $X_{H}$ are independent. So Bob's
received signal using (\ref{eq:bobs-state}) and (\ref{eq:all-noise})\ is
$\sqrt{\eta}\left(  -1\right)  ^{S+1}\alpha_{x}+N$. Bob thresholds the result
of the non-ideal homodyne detection with\ a threshold $\theta$ to retrieve a
random bit $Y$ where%
\begin{equation}
Y\equiv u\left(  \sqrt{\eta}\left(  -1\right)  ^{S+1}\alpha_{x}+N-\theta
\right)
\end{equation}
and $u$ is the unit Heaviside step function defined as $u\left(  x\right)  =1$
if $x\geq0$ and $u\left(  x\right)  =0$ if $x<0$. This final bit $Y$ that Bob
detects should be the message bit $S$ that Alice first sent.

\textit{Quantum Alpha-Stable Noise}---The noise random variable $\nu_{x}$ need
not have a finite second moment or finite higher-order moments. Some
researchers argue that quantum-optical noise arises from a large number of
independent random effects and and that it is Gaussian because of the central
limit theorem \cite{PhysRevA.50.3295,nobel2005glauber}. But these random
effects need not converge to a Gaussian random variable even though they
converge to a random variable with a bell-curve density. The
\textit{generalized} central limit theorem states that all and only normalized
stable random variables converge in distribution to a \textit{stable} random
variable \cite{book1968breiman}. So an impulsive quantum noise source may have
a limiting alpha-stable density through aggregation or directly through
transformation as when the Cauchy density arises from the tangent of uniform noise.

Alpha-stable noise models diverse physical phenomena such as impulsive
interrupts in phone lines, underwater acoustics, low-frequency atmospheric
signals, and gravitational fluctuations \cite{book1995nikias}. The parameter
$\alpha$ (different from \textquotedblleft coherent state\textquotedblright%
\ $\alpha$)\ lies in $\left(  0,2\right]  $ and parametrizes the thickness of
the curve's tails. The curve's tail thickness increases as $\alpha$ decreases:
$\alpha=1$ corresponds to the thick-tailed Cauchy random variable and
$\alpha=2$ corresponds to the familiar thin-tailed Gaussian random variable.
Parameter $\beta$ is a skewness parameter such that $\beta=0$ gives a
symmetric density. Parameter $\gamma$ is a dispersion parameter that acts like
the variance because it quantifies the spread or width of the alpha-stable
density around its location parameter $a$.

\textit{Quantum Forbidden-Interval Theorem}---Theorem~\ref{thm:QFIT} below
shows that any finite-variance quantum noise or any infinite-variance
alpha-stable noise produces the SR\ effect in our model. The theorem states
that the SR\ effect occurs for finite-variance noise if and only if the noise
mean $\mu_{\nu_{x}}$\ falls outside the forbidden interval $\left(
\theta-\alpha_{x},\theta+\alpha_{x}\right)  $. The noise location $a$ replaces
the noise mean in the forbidden-interval condition for infinite-variance
noise. So adding noise in the form of squeezing noise, channel noise, and
detector inefficiency noise can enhance the performance of the quantum
communication system. Figure~\ref{fig:basic-QSR-model}\ shows a simulation
instance of the if-part of Theorem~\ref{thm:QFIT}.

The theorem states that the mutual information $I\left(  S,Y\right)
$\ between sender and receiver tends to zero as all noise parameters decrease
to zero. The theorem assumes that the input and output signals are
statistically dependent so that $I\left(  S,Y\right)  >0$ where $I\left(
S,Y\right)  =\sum_{s,y}p_{S,Y}\left(  s,y\right)  \log\left(  p_{S,Y}\left(
s,y\right)  /\left(  p_{S}\left(  s\right)  p_{Y}\left(  y\right)  \right)
\right)  $ \cite{book1991cover}. So the SR effect occurs because the mutual
information $I\left(  S,Y\right)  $ must increase from zero as we add noise to
the system:\ \textit{what goes down must go up}. We state the parameters for
the finite-variance case without parentheses and the parameters for the
infinite-variance case with parentheses.

\begin{theorem}
\label{thm:QFIT} Suppose the position quadrature $\nu_{x}$ of the channel
noise has finite variance $\sigma_{\nu_{x}}^{2}$ and mean $\mu_{\nu_{x}}$
(dispersion $\gamma$ and location $a$). Suppose the input signal's position
quadrature $\alpha_{x}$ is subthreshold: $\alpha_{x}<\theta$. Suppose there is
some statistical dependence between input signal $S$ and output signal $Y$ so
that the mutual information obeys $I(S,Y)>0$. Then the quantum communication
system exhibits the nonmonotone SR effect if and only if the position
quadrature of the noise mean (location)\ does not lie in the forbidden
interval: $\mu_{\nu_{x}}\notin\left(  \theta-\alpha_{x},\theta+\alpha
_{x}\right)  $ ($a\notin\left(  \theta-\alpha_{x},\theta+\alpha_{x}\right)
$). The nonmonotone SR effect is that $I(S,Y)\rightarrow0$ as $\sigma_{\nu
_{x}}^{2}\rightarrow0$ ($\gamma\rightarrow0$), as $r\rightarrow\infty$, and as
$\eta\rightarrow1$.
\end{theorem}

\begin{proof}
The finite-variance proof for sufficiency and necessity follows the respective
proofs in \cite{nn2003kosko} and \cite{pre2004kosko} if we use $p_{N}\left(
n\right)  $ as the noise density. The infinite-variance proof follows the
respective stable proofs in \cite{nn2003kosko} and \cite{pre2004kosko} if we
use $p_{N}\left(  n\right)  $ as the noise density and if $\nu_{x}$ is an
alpha-stable random variable. Only slight modifications of the proofs account
for the homodyne efficiency $\eta$. See
Appendix~\ref{sec:proof-QFIT-finite-variance}\ for the finite-variance proof
and Appendix~\ref{sec:proof-QFIT-infinite-variance}\ for the infinite-variance proof.
\end{proof}

\textit{SR\ in Continuous-Variable Quantum Key Distribution}---The SR effect
occurs in the continuous-variable quantum key distribution (CVQKD) scenario
from \cite{arxiv2000hirano,PhysRevA.67.022308}. This CVQKD\ model thresholds a
continuous parameter to establish a secret key between Alice and Bob. We
modify the form of the above forbidden-interval theorem to include the
subtleties of the CVQKD model. The resulting theorem gives necessary and
sufficient conditions for the SR\ effect in CVQKD.

The theorems have security implications for CVQKD\ with thresholding. Suppose
that $I\left(  A,B\right)  $ is the mutual information between Alice and Bob
and that $I\left(  A,E\right)  $ is the mutual information between Alice and
an attacker Eve. The SR effect influences the privacy condition $I\left(
A,B\right)  >I\left(  A,E\right)  $ \cite{ieee1993maurer}\ because it affects
$I\left(  A,B\right)  $.

We first present the model for CVQKD
from~\cite{arxiv2000hirano,PhysRevA.67.022308} without including the attacker
Eve. Alice wants to send a random secret bit $S$ to Bob. Alice randomly sends
one of four coherent states to Bob:\ $\left\{  \left\vert \alpha\right\rangle
,\left\vert i\alpha\right\rangle ,\left\vert -\alpha\right\rangle ,\left\vert
-i\alpha\right\rangle \right\}  $ where $\alpha\in\mathbb{R}^{+}$. Random bit
$S=0$ if she sends $\left\vert -\alpha\right\rangle $ or $\left\vert
-i\alpha\right\rangle $ and $S=1$ if she sends $\left\vert \alpha\right\rangle
$ or $\left\vert i\alpha\right\rangle $. Bob randomly measures the state's
position quadrature or momentum quadrature. Alice and Bob communicate
classically after Alice sends a large quantity of quantum data to Bob. They
divide the measurement results into \textquotedblleft
correct-basis\textquotedblright\ and \textquotedblleft
incorrect-basis.\textquotedblright\ The data is correct-basis if Bob measures
the position quadrature when Alice sends $\left\{  \left\vert \alpha
\right\rangle ,\left\vert -\alpha\right\rangle \right\}  $ or if Bob measures
the momentum quadrature when Alice sends $\left\{  \left\vert i\alpha
\right\rangle ,\left\vert -i\alpha\right\rangle \right\}  $. The data is
incorrect-basis if it is not correct-basis. Alice and Bob keep only
correct-basis data. Let $x\in\mathbb{R}$ be the result of Bob's measurement.
Bob sets a threshold $\theta$ and assigns a bit value $Y$ where $Y=1$ if
$x\geq\theta$, $Y=0$ if $x\leq-\theta$, and $Y=\varepsilon$ otherwise. Symbol
$\varepsilon$ represents an inconclusive result.

Our analysis below corresponds only to correct-basis data because this data is
crucial for determining the resulting performance of the protocol. We present
the analysis only for the position-quadrature basis case. The same analysis
holds for the momentum-quadrature case.

We now present a Heisenberg-picture analysis of the above model and include
strategies that the attacker Eve can employ. The first few steps begin in the
same way as the basic protocol above with Eve controlling the noisy
channel.\ Then $\hat{x}e^{-r}+\left(  -1\right)  ^{S+1}\alpha_{x}+\nu_{x}$ is
the position-quadrature observable for the state that Eve possesses. She
performs an amplifier-beamsplitter attack \cite{namiki:024301} by first
passing the state through a phase-insensitive linear amplifier with gain
$G\geq1$ \cite{PhysRev.128.2407}. She then leaks a fraction $1-\eta_{E}$ of
the state through a beamsplitter so that Bob receives the fraction $\eta_{E}$.
The Heisenberg-picture observable that corresponds to Bob's state is%
\begin{equation}
\sqrt{\eta_{E}G}\hat{x}_{s}+\sqrt{\eta_{E}\left(  G-1\right)  }\hat{x}_{E_{1}%
}+\sqrt{1-\eta_{E}}\hat{x}_{E_{2}} \label{eq:eves-state}%
\end{equation}
where $\hat{x}_{s}=\hat{x}e^{-r}+\left(  -1\right)  ^{S+1}\alpha+\nu_{x}$.
Modes $\hat{x}_{E_{1}}$ and $\hat{x}_{E_{2}}$ are vacuum modes resulting from
the amplifier and beamsplitter and correspond to zero-mean $1/2$-variance
Gaussian random variables upon measurement. Bob then measures the above
operator by non-ideal position-quadrature homodyne detection. It collapses to
the random variable $N+\sqrt{\eta_{E}\eta_{B}G}\left(  -1\right)  ^{S+1}%
\alpha$ where $N$ sums all noise terms%
\begin{multline*}
N\equiv\sqrt{\eta_{E}\eta_{B}G}\left(  Xe^{-r}+\nu_{x}\right)  +\sqrt{\eta
_{E}\eta_{B}\left(  G-1\right)  }X_{E_{1}}+\\
\sqrt{\eta_{B}\left(  1-\eta_{E}\right)  }X_{E_{2}}+\sqrt{1-\eta_{B}}X_{H},
\end{multline*}
$\eta_{B}$ is the efficiency of Bob's homodyne detection, and $X_{H}$ is a
zero-mean $1/2$-variance Gaussian random variable that arises from homodyne
detection noise. The density $p_{N}\left(  n\right)  $ of random variable $N$
is%
\begin{equation}
p_{N}\left(  n\right)  =\left(  p_{\mathcal{N}\left(  0,\sigma^{2}\right)
}\ast p_{\sqrt{\eta_{E}\eta_{B}G}\nu_{x}}\right)  \left(  n\right)
\label{eq:cvqkd-density}%
\end{equation}
where $p_{\mathcal{N}\left(  0,\sigma^{2}\right)  }$ is the density of a
zero-mean Gaussian random variable with variance%
\[
\left(  \eta_{B}\left(  \eta_{E}Ge^{-2r}+\eta_{E}\left(  G-1\right)  +\left(
1-\eta_{E}\right)  \right)  +1-\eta_{B}\right)  /2
\]
and $p_{\sqrt{\eta_{E}\eta_{B}G}\nu_{x}}$ is the density of $\sqrt{\eta
_{E}\eta_{B}G}\nu_{x}$. Bob decodes with a threshold $\theta$ and gets a
random bit $Y$ where%
\begin{equation}
Y=\left\{
\begin{array}
[c]{lll}%
1 & : & N+\sqrt{\eta_{E}\eta_{B}G}\left(  -1\right)  ^{S+1}\alpha\geq\theta\\
0 & : & N+\sqrt{\eta_{E}\eta_{B}G}\left(  -1\right)  ^{S+1}\alpha\leq-\theta\\
\varepsilon & : & \text{else}%
\end{array}
\right.  .
\end{equation}

\begin{figure}
[ptb]
\begin{center}
\includegraphics[
height=2.175in,
width=2.8885in
]%
{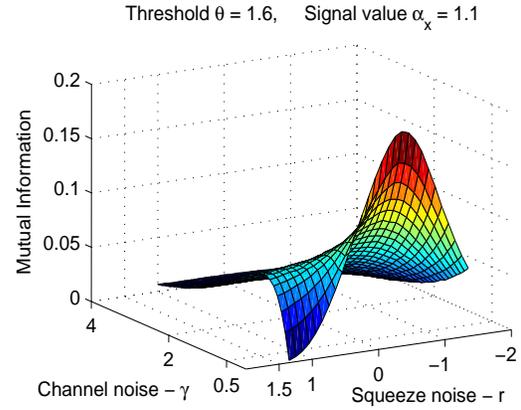}%
\caption{(Color online) SR in continuous-variable quantum key distribution.
Alice encodes coherent states with amplitude $A=1.1$ and Bob decodes with
threshold $\theta=1.6$. The graph shows the smoothed mutual information as a
function of the dispersion $\gamma$ of infinite-variance quantum Cauchy noise
and squeezing strength $r$ for 100 simulation runs. We do not include
amplifier, beamsplitter, or photodetector inefficiency noise. Each run
generated 10,000 input-output signal pairs to estimate the mutual information.
The SR effect occurs because the channel noise location $a=0$ and so $a$ lies
outside the forbidden interval $(-2.7,-.5)\cup(.5,2.7)$.}%
\label{fig:CVQKD}%
\end{center}
\end{figure}

Protagonists Alice and Bob and antagonist Eve all play a role in the
SR\ effect in Alice and Bob's communication of a secret key. Alice can add
Heisenberg noise in the form of squeezing. Eve can add channel, amplifier, and
leakage noise in her attack. Bob can add photodetector inefficiency noise. The
modified quantum forbidden-interval theorem characterizes this interplay and
gives a necessary and sufficient condition for the SR effect.
Figure~\ref{fig:CVQKD}\ shows a simulation instance of the if-part of the theorem.%

\begin{theorem}
\label{thm:qkd-finite-var-QFIT}Suppose the channel noise position quadrature
has finite variance $\sigma_{\nu_{x}}^{2}$ and mean $\mu_{\nu_{x}}$
(dispersion $\gamma$ and location $a$). Suppose the input signal's amplitude
$\alpha$ is subthreshold: $\alpha<\theta$ and $-\alpha>-\theta$. Suppose there
is some statistical dependence between input signal $S$ and output signal $Y$
so that the mutual information obeys $I(S,Y)>0$. Then the quantum key
distribution system exhibits the nonmonotone SR effect if and only if the
position quadrature of the noise mean (location) does not lie in the forbidden
interval: $\mu_{\nu_{x}}\notin\left(  -\theta-\alpha,-\theta+\alpha\right)
\cup\left(  \theta-\alpha,\theta+\alpha\right)  $ ($a\notin\left(
-\theta-\alpha,-\theta+\alpha\right)  \cup\left(  \theta-\alpha,\theta
+\alpha\right)  $). The nonmonotone SR effect is that $I(S,Y)\rightarrow0$ as
$\sigma_{\nu_{x}}^{2}\rightarrow0$ ($\gamma\rightarrow0$), as $r\rightarrow
\infty$, as $G\rightarrow1$, as $\eta_{E}\rightarrow1$, and as $\eta
_{B}\rightarrow1$.
\end{theorem}

\begin{proof}
The proof method follows the proof of Theorem~\ref{thm:QFIT} using
$p_{N}\left(  n\right)  $ in (\ref{eq:cvqkd-density}). The proof requires
three cases rather than two because of the differences between CVQKD\ and the
basic model. See Appendix~\ref{sec:proof-CVQKD-QFIT-finite-variance}\ for the
finite-variance proof and
Appendix~\ref{sec:proof-CVQKD-QFIT-infinite-variance}\ for the
infinite-variance proof.
\end{proof}

\textit{Conclusion}---Theorems~\ref{thm:QFIT} and
\ref{thm:qkd-finite-var-QFIT}\ guarantee only that the nonmonotone SR effect
occurs. They do not give the optimal combination of channel noise, squeezing,
and photodetector inefficiency noise. Nor do they guarantee a large increase
in mutual information. The theorems also may not appear realistic because
their proof requires infinite squeezing in the limit and thus requires
infinite energy. But the theorems guarantee that the SR effect occurs for some
finite squeezing. The simulations in both figures display the full nonmonotone
SR signature for experimentally plausible squeezing values and for realistic
channel noise levels.

Forbidden interval theorems may hold for more complex quantum systems. The
quantum systems in this paper use noisy quantum processing to produce a
mutual-information benefit between two classical variables. Other systems
might use noise to enhance the fidelity of the coherent superposition of a
quantum state. The performance measure would be the coherent information
\cite{book2000mikeandike} because it corresponds operationally to the capacity
of a quantum channel \cite{ieee2005dev}. The coherent information also relates
to the quantum channel capacity for sending private classical information
\cite{ieee2005dev}. This suggests further connections between SR and QKD and
the potential for new learning algorithms that can locate any noise optima.

The authors thank Todd A. Brun, Igor Devetak, Jonathan P. Dowling, and Austin
Lund for helpful discussions. MMW\ acknowledges support from NSF Grant
CCF-0545845, the Hearne Institute for Theoretical Physics, Army Research
Office, and Disruptive Technologies Office.

\section{Appendix}

\subsection{Proof of Theorem \ref{thm:QFIT} (Finite Variance)}

\label{sec:proof-QFIT-finite-variance}The proofs for sufficiency and necessity
follow the respective proof methods in \cite{nn2003kosko} and
\cite{pre2004kosko} if we use (\ref{eq:noise-density}) as the noise density.

Let us calculate the four conditional probabilities $p_{Y|S}(0|0)$,
$p_{Y|S}(0|1)$, $p_{Y|S}(1|0)$, $p_{Y|S}(1|1)$.%
\begin{align}
&  p_{Y|S}(0|0)\nonumber\\
&  =\Pr\left\{  u\left(  \sqrt{\eta}\left(  -1\right)  ^{S+1}\alpha
_{x}+N-\theta\right)  \ |\ S=0\right\} \nonumber\\
&  =\Pr\left\{  \sqrt{\eta}\left(  -1\right)  ^{S+1}\alpha_{x}+N-\theta
<0\ |\ S=0\right\} \nonumber\\
&  =\Pr\left\{  -\sqrt{\eta}\alpha_{x}+N<\theta\right\}  =\Pr\left\{
N<\theta+\sqrt{\eta}\alpha_{x}\right\} \nonumber\\
&  =\int_{-\infty}^{\theta+\sqrt{\eta}\alpha_{x}}p_{N}\left(  n\right)  \ dn
\end{align}
The other conditional probabilities follow from similar calculations:%
\begin{align}
p_{Y|S}(0|1)  &  =\int_{-\infty}^{\theta-\sqrt{\eta}\alpha_{x}}p_{N}\left(
n\right)  \ dn\\
p_{Y|S}(1|0)  &  =\int_{\theta+\sqrt{\eta}\alpha_{x}}^{\infty}p_{N}\left(
n\right)  \ dn\\
p_{Y|S}(1|1)  &  =\int_{\theta-\sqrt{\eta}\alpha_{x}}^{\infty}p_{N}\left(
n\right)  \ dn
\end{align}

\begin{proof}
[Proof (Sufficiency)]Assume that $0<p_{S}\left(  s\right)  <1$ to avoid
triviality when $p_{S}\left(  s\right)  =0$ or $1$. $I\left(  S,Y\right)  =0$
if and only if $S$ and $Y$ are statistically independent \cite{book1991cover}.
We show that $S$ and $Y$ are asymptotically independent:\ $I\left(
S,Y\right)  \rightarrow0$ as $\sigma_{\nu_{x}}^{2}\rightarrow0$, as
$r\rightarrow\infty$, and as $\eta\rightarrow1$. The following definition
holds for the proofs that follow:%
\begin{equation}
\sigma^{2}\equiv\eta\left(  e^{-2r}/2+\sigma_{\nu_{x}}^{2}\right)  +\left(
1-\eta\right)  /2
\end{equation}
We need to show that $p_{Y|S}(y|s)=p_{Y}(y)$ as $\sigma_{\nu_{x}}%
^{2}\rightarrow0$, as $r\rightarrow\infty$, and as $\eta\rightarrow1$ for
$s,y\in\left\{  0,1\right\}  $. Consider an algebraic manipulation using the
law of total probability:%
\begin{align}
p_{Y}(y) &  =\sum_{s}p_{Y|S}(y|s)\ p_{S}(s)\label{eq:law_tot_prob}\\
&  =p_{Y|S}(y|0)\ p_{S}(0)+p_{Y|S}(y|1)\ p_{S}(1)\nonumber\\
&  =p_{Y|S}(y|0)\ p_{S}(0)+p_{Y|S}(y|1)\ \left(  1-p_{S}(0)\right)
\nonumber\\
&  =\left(  p_{Y|S}(y|0)-p_{Y|S}(y|1)\right)  p_{S}(0)+p_{Y|S}(y|1)\nonumber
\end{align}
We can show by a similar method that
\[
p_{Y}(y)=\left(  p_{Y|S}(y|1)-p_{Y|S}(y|0)\right)  p_{S}(1)+p_{Y|S}(y|0)
\]
So $p_{Y}(y)\rightarrow p_{Y|S}(y|1)$ and $p_{Y}(y)\rightarrow p_{Y|S}(y|0)$
as $p_{Y|S}(y|1)-p_{Y|S}(y|0)\rightarrow0$. Consider the case where $y=0$.%
\[
p_{Y|S}(0|0)-p_{Y|S}(0|1)=\int_{\theta-\sqrt{\eta}\alpha_{x}}^{\theta
+\sqrt{\eta}\alpha_{x}}p_{N}\left(  n\right)  \ dn
\]
Consider the case where $y=1$.%
\[
p_{Y|S}(1|0)-p_{Y|S}(1|1)=-\int_{\theta-\sqrt{\eta}\alpha_{x}}^{\theta
+\sqrt{\eta}\alpha_{x}}p_{N}\left(  n\right)  \ dn
\]
So the result follows if
\begin{equation}
\int_{\theta-\sqrt{\eta}\alpha_{x}}^{\theta+\sqrt{\eta}\alpha_{x}}p_{N}\left(
n\right)  \ dn\rightarrow0
\end{equation}
as $\sigma_{\nu_{x}}^{2}\rightarrow0$, as $r\rightarrow\infty$, and as
$\eta\rightarrow1$. Suppose the mean
\[
\mu_{\nu_{x}}\notin\left(  \theta-\alpha_{x},\theta+\alpha_{x}\right)
\]
by hypothesis. We ignore the zero-measure cases where $\mu_{\nu_{x}}%
=\theta-\alpha_{x}$ or $\mu_{\nu_{x}}=\theta+\alpha_{x}$. \newline\newline
Case 1:\ Suppose first that $\mu_{\nu_{x}}<$ $\theta-\alpha_{x}$. So $\mu
_{\nu_{x}}+\alpha_{x}<$ $\theta$ and thus
\[
\sqrt{\eta}\left(  \mu_{\nu_{x}}+\alpha_{x}\right)  \leq\mu_{\nu_{x}}%
+\alpha_{x}<\theta
\]
for any $\eta\in\left(  0,1\right]  $. Pick
\[
\epsilon=\frac{1}{2}\left(  \theta-\sqrt{\eta}\alpha_{x}-\sqrt{\eta}\mu
_{\nu_{x}}\right)  >0.
\]
So $\theta-\sqrt{\eta}\alpha_{x}-\epsilon=\sqrt{\eta}\mu_{x}+\epsilon.$ Then%
\begin{align*}
&  \int_{\theta-\sqrt{\eta}\alpha_{x}}^{\theta+\sqrt{\eta}\alpha_{x}}%
p_{N}\left(  n\right)  \ dn\\
&  \leq\int_{\theta-\sqrt{\eta}\alpha_{x}}^{\infty}p_{N}\left(  n\right)
\ dn\\
&  \leq\int_{\theta-\sqrt{\eta}\alpha_{x}-\epsilon}^{\infty}p_{N}\left(
n\right)  \ dn\\
&  \leq\int_{\sqrt{\eta}\mu_{\nu_{x}}+\epsilon}^{\infty}p_{N}\left(  n\right)
\ dn\\
&  =\text{Pr}\left\{  N\geq\sqrt{\eta}\mu_{\nu_{x}}+\epsilon\right\}  \\
&  =\text{Pr}\left\{  N\geq\mu+\epsilon\right\}  \\
&  =\text{Pr}\left\{  N-\mu\geq\epsilon\right\}  \\
&  \leq\text{Pr}\left\{  \left\vert N-\mu\right\vert \geq\epsilon\right\}  \\
&  \leq\frac{\sigma^{2}}{\epsilon^{2}}%
\end{align*}
So the result follows when $\mu_{\nu_{x}}<$ $\theta-\alpha_{x}$ because
$p_{Y|S}(0|0)-p_{Y|S}(0|1)\rightarrow0$ as $\sigma_{\nu_{x}}^{2}\rightarrow0$,
as $r\rightarrow\infty$, and as $\eta\rightarrow1$.\newline\newline Case 2:
Suppose next that $\mu_{\nu_{x}}>\theta+\alpha_{x}$ so that $\mu_{\nu_{x}%
}-\alpha_{x}>\theta>0$. Choose $\sqrt{\eta}$ large enough so that
\[
\sqrt{\eta}>\theta/\left(  \mu_{\nu_{x}}-\alpha_{x}\right)
\]
So $\sqrt{\eta}\left(  \mu_{\nu_{x}}-\alpha_{x}\right)  >\theta$. Pick
\[
\epsilon=\frac{1}{2}\left(  \sqrt{\eta}\mu_{\nu_{x}}-\theta-\sqrt{\eta}%
\alpha_{x}\right)  >0.
\]
So $\theta+\sqrt{\eta}\alpha_{x}+\epsilon=\sqrt{\eta}\mu_{\nu_{x}}-\epsilon.$
Then%
\begin{align*}
&  \int_{\theta-\sqrt{\eta}\alpha_{x}}^{\theta+\sqrt{\eta}\alpha_{x}}%
p_{N}\left(  n\right)  \ dn\\
&  \leq\int_{-\infty}^{\theta+\sqrt{\eta}\alpha_{x}}p_{N}\left(  n\right)
\ dn\\
&  \leq\int_{-\infty}^{\theta+\sqrt{\eta}\alpha_{x}+\epsilon}p_{N}\left(
n\right)  \ dn\\
&  \leq\int_{-\infty}^{\sqrt{\eta}\mu_{\nu_{x}}-\epsilon}p_{N}\left(
n\right)  \ dn\\
&  =\text{Pr}\left\{  N\leq\sqrt{\eta}\mu_{\nu_{x}}-\epsilon\right\}  \\
&  =\text{Pr}\left\{  N\leq\mu-\epsilon\right\}  \\
&  =\text{Pr}\left\{  N-\mu\leq-\epsilon\right\}  \\
&  \leq\text{Pr}\left\{  \left\vert N-\mu\right\vert \geq\epsilon\right\}  \\
&  \leq\frac{\sigma^{2}}{\epsilon^{2}}%
\end{align*}
So $p_{Y|S}(0|0)-p_{Y|S}(0|1)\rightarrow0$ as $\sigma_{\nu_{x}}^{2}%
\rightarrow0$, as $r\rightarrow\infty$, and as $\eta\rightarrow1$ when
$\mu_{\nu_{x}}>\theta+\alpha_{x}$. Thus
\[
\mu_{\nu_{x}}\notin\left(  \theta-\alpha_{x},\theta+\alpha_{x}\right)
\]
is a sufficient condition for the nonmonotone SR effect to occur.
\end{proof}

\begin{proof}
[Proof (Necessity)]The system does not exhibit the nonmonotone SR\ effect if
$\mu_{\nu_{x}}\in\left(  \theta-\alpha_{x},\theta+\alpha_{x}\right)  $ in the
sense that $I(S,Y)$ is maximum as $\sigma_{\nu_{x}}^{2}\rightarrow0$, as
$r\rightarrow\infty$, and as $\eta\rightarrow1$. $I(S,Y)\rightarrow H\left(
Y\right)  =H\left(  S\right)  $ as $\sigma_{\nu_{x}}^{2}\rightarrow0$, as
$r\rightarrow\infty$, and as $\eta\rightarrow1$. Assume that $0<p_{S}\left(
s\right)  <1$ to avoid triviality when $p_{S}\left(  s\right)  =0$ or $1$. We
show that $H\left(  Y\right)  \rightarrow H\left(  S\right)  $ and $H\left(
Y|S\right)  \rightarrow0$ as $\sigma_{\nu_{x}}^{2}\rightarrow0$, as
$r\rightarrow\infty$, and as $\eta\rightarrow1$. It is maximum in this limit
because $I(S,Y)=H\left(  Y\right)  -H\left(  Y|S\right)  $ and $I(S,Y)\leq
H\left(  S\right)  $ by the data processing inequality for a Markov chain
\cite{book1991cover}. Consider the conditional entropy $H\left(  Y|S\right)
$:%
\begin{align}
&  H\left(  Y|S\right)  \nonumber\\
&  =-\sum_{s,y}p_{Y,S}\left(  y,s\right)  \log_{2}p_{Y|S}\left(  y|s\right)
\nonumber\\
&  =-\sum_{s}p_{S}\left(  s\right)  \sum_{y}p_{Y|S}\left(  y|s\right)
\log_{2}p_{Y|S}\left(  y|s\right)  \label{eq:cond-ent}%
\end{align}
Suppose for now that $p_{Y|S}(y|s)\rightarrow1$ or $0$ for all $s,y\in\left\{
0,1\right\}  $ as $\sigma_{\nu_{x}}^{2}\rightarrow0$, as $r\rightarrow\infty$,
and as $\eta\rightarrow1$. Then $H\left(  Y|S\right)  \rightarrow0$ by
inspecting (\ref{eq:cond-ent}) and applying $1\log_{2}1=0$ and $0\log_{2}0=0$
by L'H\^{o}spital's rule. So we aim to prove that each of the conditional
probabilities vanish or approach $1$ in the above limit if $\mu_{\nu_{x}}%
\in\left(  \theta-\alpha_{x},\theta+\alpha_{x}\right)  $. Consider first
$p_{Y|S}(0|0)$. Pick any $\mu_{\nu_{x}}\in\left(  \theta-\alpha_{x}%
,\theta+\alpha_{x}\right)  $. Then $\theta+\alpha_{x}-\mu_{\nu_{x}}>0$ and
$\theta>\mu_{\nu_{x}}-\alpha_{x}$. Then $\theta>\sqrt{\eta}\left(  \mu
_{\nu_{x}}-\alpha_{x}\right)  $ for any $\eta\in\left(  0,1\right]  $. Pick
$\epsilon=\frac{1}{2}\left(  \theta+\sqrt{\eta}\alpha_{x}-\sqrt{\eta}\mu
_{\nu_{x}}\right)  >0$ so that $\theta+\sqrt{\eta}\alpha_{x}-\epsilon
=\sqrt{\eta}\mu_{\nu_{x}}+\epsilon$.%
\begin{align*}
&  p_{Y|S}(0|0)\\
&  =\int_{-\infty}^{\theta+\sqrt{\eta}\alpha_{x}}p_{N}\left(  n\right)  \ dn\\
&  \geq\int_{-\infty}^{\theta+\sqrt{\eta}\alpha_{x}-\epsilon}p_{N}\left(
n\right)  \ dn\\
&  =\int_{-\infty}^{\sqrt{\eta}\mu_{\nu_{x}}+\epsilon}p_{N}\left(  n\right)
\ dn\\
&  =1-\int_{\sqrt{\eta}\mu_{\nu_{x}}+\epsilon}^{\infty}p_{N}\left(  n\right)
\ dn\\
&  =1-\text{Pr}\left\{  N\geq\sqrt{\eta}\mu_{\nu_{x}}+\epsilon\right\}  \\
&  =1-\text{Pr}\left\{  N\geq\mu+\epsilon\right\}  \\
&  =1-\text{Pr}\left\{  N-\mu\geq\epsilon\right\}  \\
&  \geq1-\text{Pr}\left\{  \left\vert N-\mu\right\vert \geq\epsilon\right\}
\\
&  \geq1-\frac{\sigma^{2}}{\epsilon^{2}}\\
&  \rightarrow1
\end{align*}
as $\sigma_{\nu_{x}}^{2}\rightarrow0,$as $r\rightarrow\infty,$and as
$\eta\rightarrow1$. We prove the result similarly for $p_{Y|S}(1|1)$. Pick any
$\mu_{\nu_{x}}\in\left(  \theta-\alpha_{x},\theta+\alpha_{x}\right)  $. Then
$\mu_{\nu_{x}}>\theta-\alpha_{x}$ and $\mu_{\nu_{x}}+\alpha_{x}>\theta>0$.
Suppose that $\eta$ is large enough so that $\sqrt{\eta}>\theta/\left(
\mu_{\nu_{x}}+\alpha_{x}\right)  $. Then $\sqrt{\eta}\left(  \mu_{\nu_{x}%
}+\alpha_{x}\right)  >\theta$ and $\sqrt{\eta}\mu_{\nu_{x}}+\sqrt{\eta}%
\alpha_{x}-\theta>0$. Pick $\epsilon=\frac{1}{2}\left(  \sqrt{\eta}\mu
_{\nu_{x}}+\sqrt{\eta}\alpha_{x}-\theta\right)  >0$ so that $\theta-\sqrt
{\eta}\alpha_{x}+\epsilon=\sqrt{\eta}\mu_{\nu_{x}}-\epsilon$.%
\begin{align*}
&  p_{Y|S}(1|1)\\
&  =\int_{\theta-\sqrt{\eta}\alpha_{x}}^{\infty}p_{N}\left(  n\right)  \ dn\\
&  \geq\int_{\theta-\sqrt{\eta}\alpha_{x}+\epsilon}^{\infty}p_{N}\left(
n\right)  \ dn\\
&  =\int_{\sqrt{\eta}\mu_{\nu_{x}}-\epsilon}^{\infty}p_{N}\left(  n\right)
\ dn\\
&  =1-\int_{-\infty}^{\sqrt{\eta}\mu_{\nu_{x}}-\epsilon}p_{N}\left(  n\right)
\ dn\\
&  =1-\text{Pr}\left\{  N\leq\sqrt{\eta}\mu_{\nu_{x}}-\epsilon\right\}  \\
&  =1-\text{Pr}\left\{  N\leq\mu-\epsilon\right\}  \\
&  =1-\text{Pr}\left\{  N-\mu\leq-\epsilon\right\}  \\
&  \geq1-\text{Pr}\left\{  \left\vert N-\mu\right\vert \geq\epsilon\right\}
\\
&  \geq1-\frac{\sigma^{2}}{\epsilon^{2}}\\
&  \rightarrow1
\end{align*}
as $\sigma_{\nu_{x}}^{2}\rightarrow0,$as $r\rightarrow\infty,$and as
$\eta\rightarrow1$. So $p_{Y|S}(0|0)\rightarrow1,p_{Y|S}(1|1)\rightarrow
1,p_{Y|S}(1|0)\rightarrow0,$ and $p_{Y|S}(0|1)\rightarrow0$ as $\sigma
_{\nu_{x}}^{2}\rightarrow0$, as $r\rightarrow\infty$, and as $\eta
\rightarrow1$ and the system does not display the nonmonotone
SR\ effect.\pagebreak
\end{proof}

\subsection{Proof of Theorem \ref{thm:QFIT} (Infinite Variance)}

\label{sec:proof-QFIT-infinite-variance}The proofs for sufficiency and
necessity follow the respective stable proof methods in \cite{nn2003kosko} and
\cite{pre2004kosko} if we use (\ref{eq:noise-density}) as the noise density
and if $\nu_{x}$ is an alpha-stable random variable.

The characteristic function $\varphi_{\nu_{x}}\left(  \omega\right)  $ of an
alpha-stable noise source with density $p_{\nu_{x}}\left(  n\right)  $ is the
following:%
\begin{multline}
\varphi_{\nu_{x}}\left(  \omega\right)
=\label{eq:characteristic-alpha-stable}\\
\exp\left\{  ia\omega-\gamma\left\vert \omega\right\vert ^{\alpha}\left(
1+i\beta\text{sign}(\omega)\tan\left(  \frac{\alpha\pi}{2}\right)  \right)
\right\}
\end{multline}
where $\alpha$ is the characteristic exponent and $\beta$ is a skewness
parameter. The characteristic function of $p_{N}\left(  n\right)  $ is as
follows%
\begin{align}
&  \varphi_{N}\left(  \omega\right) \nonumber\\
&  =\left(  \varphi_{\sqrt{\eta}Xe^{-r}}\cdot\varphi_{\sqrt{\eta}\nu_{x}}%
\cdot\varphi_{\sqrt{1-\eta}X_{H}}\right)  \left(  \omega\right) \nonumber\\
&  =\exp\left\{  -\frac{\eta e^{-2r}\omega^{2}}{4}\right\}  \varphi_{\nu_{x}%
}\left(  \sqrt{\eta}\omega\right)  \exp\left\{  -\frac{\left(  1-\eta\right)
\omega^{2}}{4}\right\} \nonumber\\
&  =\varphi_{\nu_{x}}\left(  \sqrt{\eta}\omega\right)  \exp\left\{
-\frac{\left(  \eta e^{-2r}+1-\eta\right)  \omega^{2}}{4}\right\}
\end{align}
from (\ref{eq:noise-density}) and the convolution theorem.

\begin{proof}
[Proof (Sufficiency)]Take the limit of the characteristic function
$\varphi_{N}\left(  \omega\right)  $ as $\gamma\rightarrow0$, as squeezing
parameter $r\rightarrow\infty$, and as homodyne efficiency $\eta\rightarrow1$
to obtain the following characteristic function.%
\begin{equation}
\lim_{r\rightarrow\infty,\gamma\rightarrow0,\eta\rightarrow1}\varphi
_{N}(\omega)=\exp\left\{  ia\omega\right\}
\end{equation}
The probability density $p_{N}\left(  n\right)  $ then approaches a translated
delta function%
\begin{equation}
\lim_{r\rightarrow\infty,\gamma\rightarrow0,\eta\rightarrow1}p_{N}\left(
n\right)  =\delta\left(  n-a\right)
\end{equation}
The conditional probability difference obeys:%
\begin{align}
p_{Y|S}(0|0)-p_{Y|S}(0|1) &  =\int_{\theta-\sqrt{\eta}\alpha_{x}}%
^{\theta+\sqrt{\eta}\alpha_{x}}p_{N}\left(  n\right)  \ dn\\
&  \leq\int_{\theta-\alpha_{x}}^{\theta+\alpha_{x}}p_{N}\left(  n\right)  \ dn
\end{align}
Pick $a\notin\left(  \theta-\alpha_{x},\theta+\alpha_{x}\right)  $. Consider
the following limit:%
\begin{align}
&  \lim_{r\rightarrow\infty,\gamma\rightarrow0,\eta\rightarrow1}%
p_{Y|S}(0|0)-p_{Y|S}(0|1)\\
&  \leq\lim_{r\rightarrow\infty,\gamma\rightarrow0,\eta\rightarrow1}%
\int_{\theta-\alpha_{x}}^{\theta+\alpha_{x}}p_{N}\left(  n\right)  \ dn\\
&  =\int_{\theta-\alpha_{x}}^{\theta+\alpha_{x}}\delta\left(  n-a\right)
\ dn=0
\end{align}
because $a\notin\left(  \theta-\alpha_{x},\theta+\alpha_{x}\right)  $.
\end{proof}

\begin{proof}
[Proof (Necessity)]Choose $a\in\left(  \theta-\alpha_{x},\theta+\alpha
_{x}\right)  $. Then%
\begin{align}
p_{Y|S}(0|0) &  =\int_{-\infty}^{\theta+\sqrt{\eta}\alpha_{x}}p_{N}\left(
n\right)  \ dn\\
&  =\int_{-\infty}^{\theta}p_{N}\left(  n+\sqrt{\eta}\alpha_{x}\right)  \ dn\\
&  \rightarrow\int_{-\infty}^{\theta}\delta\left(  n-a+\alpha_{x}\right)
\ dn\\
&  =\int_{-\infty}^{\theta+\alpha_{x}}\delta\left(  n-a\right)  \ dn=1\\
\text{as }\gamma &  \rightarrow0\text{, as }r\rightarrow\infty\text{, and as
}\eta\rightarrow1\\
p_{Y|S}(1|1) &  =\int_{\theta-\sqrt{\eta}\alpha_{x}}^{\infty}p_{N}\left(
n\right)  \ dn\\
&  =\int_{\theta}^{\infty}p_{N}\left(  n-\sqrt{\eta}\alpha_{x}\right)  \ dn\\
&  \rightarrow\int_{\theta}^{\infty}\delta\left(  n-a_{x}-\alpha_{x}\right)
\ dn\\
&  =\int_{\theta-\alpha_{x}}^{\infty}\delta\left(  n-a_{x}\right)  \ dn=1\\
\text{as }\gamma &  \rightarrow0\text{, as }r\rightarrow\infty\text{, and as
}\eta\rightarrow1
\end{align}
\pagebreak
\end{proof}

\subsection{Proof of Theorem \ref{thm:qkd-finite-var-QFIT} (Finite Variance)}

\label{sec:proof-CVQKD-QFIT-finite-variance}The mean $\mu$\ and variance
$\sigma^{2}$ of noise random variable $N$ are $\mu=\sqrt{\eta_{E}\eta_{B}G}%
\mu_{\nu_{x}}$ and%
\begin{multline}
\sigma^{2}=\eta_{E}\eta_{B}G\sigma_{\nu_{x}}^{2}+\label{eq:CVQKD-var}\\
\left(  \eta_{B}\left(
\begin{array}
[c]{c}%
\eta_{E}Ge^{-2r}+\eta_{E}\left(  G-1\right)  \\
+\left(  1-\eta_{E}\right)
\end{array}
\right)  +1-\eta_{B}\right)  /2.
\end{multline}
We compute the six conditional probabilities: $p_{Y|S}(0|0)$, $p_{Y|S}(0|1)$,
$p_{Y|S}(1|0)$, $p_{Y|S}(1|1)$, $p_{Y|S}(\varepsilon|0)$, and $p_{Y|S}%
(\varepsilon|1)$.
\begin{align}
&  p_{Y|S}(0|0)\nonumber\\
&  =\Pr\left\{  N+\sqrt{\eta_{E}\eta_{B}G}\left(  -1\right)  ^{S+1}\alpha
\leq-\theta\ |\ S=0\right\}  \nonumber\\
&  =\Pr\left\{  -\sqrt{\eta_{E}\eta_{B}G}\alpha+N\leq-\theta\right\}
\nonumber\\
&  =\Pr\left\{  N<-\theta+\sqrt{\eta_{E}\eta_{B}G}\alpha\right\}  \nonumber\\
&  =\int_{-\infty}^{-\theta+\sqrt{\eta_{E}\eta_{B}G}\alpha}p_{N}\left(
n\right)  \ dn
\end{align}
The other conditional probabilities follow from similar reasoning:%
\begin{align}
p_{Y|S}(0|1) &  =\int_{-\infty}^{-\theta-\sqrt{\eta_{E}\eta_{B}G}\alpha}%
p_{N}\left(  n\right)  \ dn\\
p_{Y|S}(1|0) &  =\int_{\theta+\sqrt{\eta_{E}\eta_{B}G}\alpha}^{\infty}%
p_{N}\left(  n\right)  \ dn\\
p_{Y|S}(1|1) &  =\int_{\theta-\sqrt{\eta_{E}\eta_{B}G}\alpha}^{\infty}%
p_{N}\left(  n\right)  \ dn
\end{align}%
\begin{align}
p_{Y|S}(\varepsilon|0) &  =1-p_{Y|S}(0|0)-p_{Y|S}(1|0)\nonumber\\
&  =\int_{-\theta+\sqrt{\eta_{E}\eta_{B}G}\alpha}^{\theta+\sqrt{\eta_{E}%
\eta_{B}G}\alpha}p_{N}\left(  n\right)  \ dn
\end{align}%
\begin{align}
p_{Y|S}(\varepsilon|1) &  =1-p_{Y|S}(0|1)-p_{Y|S}(1|1)\nonumber\\
&  =\int_{-\theta-\sqrt{\eta_{E}\eta_{B}G}\alpha}^{\theta-\sqrt{\eta_{E}%
\eta_{B}G}\alpha}p_{N}\left(  n\right)  \ dn
\end{align}

\begin{proof}
[Proof (Sufficiency)]We follow the proof method of Theorem \ref{thm:QFIT} with
some modifications. Note that the conditions $\eta_{E},\eta_{B}\leq1$ and
$G\geq1$ constrain the way in which we take both of their limits to one. We
give the constraint on the values that the root of their product $\sqrt
{\eta_{E}\eta_{B}G}$ may take for any given value of the noise mean $\mu
_{\nu_{x}}$. Assume that these constraints are implicit when considering the
limit in the proofs that follow.\newline\newline Assume that $0<p_{S}\left(
s\right)  <1$ to avoid triviality when $p_{S}\left(  s\right)  =0$ or $1$.
$I\left(  S,Y\right)  =0$ if and only if $S$ and $Y$ are statistically
independent \cite{book1991cover}. We show that $S$ and $Y$ are asymptotically
independent:\ $I\left(  S,Y\right)  \rightarrow0$ as $\sigma_{\nu_{x}}%
^{2}\rightarrow0$, as $r\rightarrow\infty$, as $\eta\rightarrow1$, and as
$G\rightarrow1$. We need to show that $p_{Y|S}(y|s)\rightarrow p_{Y}(y)$ as
$\sigma_{\nu_{x}}^{2}\rightarrow0$, as $r\rightarrow\infty$, as $\eta
\rightarrow1$, and as $G\rightarrow1$ for $s,y\in\left\{  0,1\right\}  $. We
do not consider $p_{Y|S}(y|\varepsilon)$ because the probability $p_{S}\left(
\varepsilon\right)  $ is zero and so the probability $p_{Y}\left(
\varepsilon\right)  $ is also zero. Consider the expansion in
(\ref{eq:law_tot_prob}) using the law of total probability. The expansion is
the same even when including symbol $\varepsilon$ because $\varepsilon$ has
zero probability:\ $p_{S}\left(  \varepsilon\right)  =0$. So $p_{Y}%
(y)\rightarrow p_{Y|S}(y|1)$ and $p_{Y}(y)\rightarrow p_{Y|S}(y|0)$ as
$p_{Y|S}(0|0)-p_{Y|S}(0|1)\rightarrow0$ and $p_{Y|S}(1|1)-p_{Y|S}%
(1|0)\rightarrow0$. Consider the case where $y=0$.%
\begin{equation}
p_{Y|S}(0|0)-p_{Y|S}(0|1)=\int_{-\theta-\sqrt{\eta_{E}\eta_{B}G}\alpha
}^{-\theta+\sqrt{\eta_{E}\eta_{B}G}\alpha}p_{N}\left(  n\right)
\ dn\label{eq:cond-prob-diff-1}%
\end{equation}
Consider the case where $y=1$.%
\begin{equation}
p_{Y|S}(1|1)-p_{Y|S}(1|0)=\int_{\theta-\sqrt{\eta_{E}\eta_{B}G}\alpha}%
^{\theta+\sqrt{\eta_{E}\eta_{B}G}\alpha}p_{N}\left(  n\right)
\ dn\label{eq:cond-prob-diff-2}%
\end{equation}
So the result follows if both of the above conditional probability differences
vanish as $\sigma_{\nu_{x}}^{2}\rightarrow0$, as $r\rightarrow\infty$, as
$\eta_{E},\eta_{B}\rightarrow1$, and as $G\rightarrow1$. Suppose the mean
$\mu_{\nu_{x}}\notin\left(  -\theta-\alpha,-\theta+\alpha\right)  \cup\left(
\theta-\alpha,\theta+\alpha\right)  $ by hypothesis. We ignore the
zero-measure cases where $\mu_{\nu_{x}}=\theta-\alpha$, $\mu_{\nu_{x}}%
=\theta+\alpha$, $\mu_{\nu_{x}}=-\theta+\alpha$, or $\mu_{\nu_{x}}%
=-\theta-\alpha$.\newline\newline Case 1: Suppose first that $\mu_{\nu_{x}%
}<-\theta-\alpha$. So $\mu_{\nu_{x}}+\alpha<-\theta$ and thus $\sqrt{\eta
_{E}\eta_{B}G}\left(  \mu_{\nu_{x}}+\alpha\right)  \leq\mu_{\nu_{x}}%
+\alpha<-\theta$ whenever $\sqrt{\eta_{E}\eta_{B}G}>-\theta/\left(  \mu
_{\nu_{x}}+\alpha\right)  =\theta/\left\vert \mu_{\nu_{x}}+\alpha\right\vert
$. Pick $\epsilon=\frac{1}{2}\left(  -\theta-\sqrt{\eta_{E}\eta_{B}G}%
\alpha-\sqrt{\eta_{E}\eta_{B}G}\mu_{\nu_{x}}\right)  >0$. So $-\theta
-\sqrt{\eta_{E}\eta_{B}G}\alpha-\epsilon=\sqrt{\eta_{E}\eta_{B}G}\mu_{\nu_{x}%
}+\epsilon$. Then%
\begin{align*}
&  \int_{-\theta-\sqrt{\eta_{E}\eta_{B}G}\alpha}^{-\theta+\sqrt{\eta_{E}%
\eta_{B}G}\alpha}p_{N}\left(  n\right)  \ dn\\
&  \leq\int_{-\theta-\sqrt{\eta_{E}\eta_{B}G}\alpha}^{\infty}p_{N}\left(
n\right)  \ dn\\
&  \leq\int_{-\theta-\sqrt{\eta_{E}\eta_{B}G}\alpha-\epsilon}^{\infty}%
p_{N}\left(  n\right)  \ dn\\
&  \leq\int_{\sqrt{\eta_{E}\eta_{B}G}\mu_{\nu_{x}}+\epsilon}^{\infty}%
p_{N}\left(  n\right)  \ dn\\
&  =\text{Pr}\left\{  N\geq\sqrt{\eta_{E}\eta_{B}G}\mu_{\nu_{x}}%
+\epsilon\right\}  \\
&  =\text{Pr}\left\{  N\geq\mu+\epsilon\right\}  \\
&  =\text{Pr}\left\{  N-\mu\geq\epsilon\right\}  \\
&  \leq\text{Pr}\left\{  \left\vert N-\mu\right\vert \geq\epsilon\right\}  \\
&  \leq\frac{\sigma^{2}}{\epsilon^{2}}%
\end{align*}
So the conditional probability difference in (\ref{eq:cond-prob-diff-1}%
)\ vanishes as $\sigma_{\nu_{x}}^{2}\rightarrow0$, as $r\rightarrow\infty$, as
$\eta_{E},\eta_{B}\rightarrow1$, and as $G\rightarrow1$ when $\mu_{\nu_{x}%
}<-\theta-\alpha$. We now prove that the conditional probability difference in
(\ref{eq:cond-prob-diff-2}) vanishes when $\mu_{\nu_{x}}<-\theta-\alpha$. It
follows that $\mu_{\nu_{x}}<\theta-\alpha$ if $\mu_{\nu_{x}}<-\theta-\alpha$.
So $\mu_{\nu_{x}}+\alpha<\theta$ and thus $\sqrt{\eta_{E}\eta_{B}G}\left(
\mu_{\nu_{x}}+\alpha\right)  \leq\mu_{\nu_{x}}+\alpha<\theta$ for any
$\sqrt{\eta_{E}\eta_{B}G}\geq0$ because $\mu_{\nu_{x}}+\alpha<0$. Pick
$\epsilon=\frac{1}{2}\left(  \theta-\sqrt{\eta_{E}\eta_{B}G}\alpha-\sqrt
{\eta_{E}\eta_{B}G}\mu_{\nu_{x}}\right)  >0$. So $\theta-\sqrt{\eta_{E}%
\eta_{B}G}\alpha-\epsilon=\sqrt{\eta_{E}\eta_{B}G}\mu_{\nu_{x}}+\epsilon$.
Then%
\begin{align*}
&  \int_{\theta-\sqrt{\eta_{E}\eta_{B}G}\alpha}^{\theta+\sqrt{\eta_{E}\eta
_{B}G}\alpha}p_{N}\left(  n\right)  \ dn\\
&  \leq\int_{\theta-\sqrt{\eta_{E}\eta_{B}G}\alpha}^{\infty}p_{N}\left(
n\right)  \ dn\\
&  \leq\int_{\theta-\sqrt{\eta_{E}\eta_{B}G}\alpha-\epsilon}^{\infty}%
p_{N}\left(  n\right)  \ dn\\
&  \leq\int_{\sqrt{\eta_{E}\eta_{B}G}\mu_{\nu_{x}}+\epsilon}^{\infty}%
p_{N}\left(  n\right)  \ dn\\
&  =\text{Pr}\left\{  N\geq\sqrt{\eta_{E}\eta_{B}G}\mu_{\nu_{x}}%
+\epsilon\right\}  \\
&  =\text{Pr}\left\{  N\geq\mu+\epsilon\right\}  \\
&  =\text{Pr}\left\{  N-\mu\geq\epsilon\right\}  \\
&  \leq\text{Pr}\left\{  \left\vert N-\mu\right\vert \geq\epsilon\right\}  \\
&  \leq\frac{\sigma^{2}}{\epsilon^{2}}%
\end{align*}
So the conditional probability difference in (\ref{eq:cond-prob-diff-2})
vanishes as $\sigma_{\nu_{x}}^{2}\rightarrow0$, as $r\rightarrow\infty$, as
$\eta_{E},\eta_{B}\rightarrow1$, and as $G\rightarrow1$ when $\mu_{\nu_{x}%
}<-\theta-\alpha$ and with constraint $\sqrt{\eta_{E}\eta_{B}G}>\theta
/\left\vert \mu_{\nu_{x}}+\alpha\right\vert $.\newline\newline Case 2: Suppose
next that $-\theta+\alpha<\mu_{\nu_{x}}<\theta-\alpha$. We first prove that
the conditional probability difference in (\ref{eq:cond-prob-diff-1}) vanishes
as $\sigma_{\nu_{x}}^{2}\rightarrow0$, as $r\rightarrow\infty$, as $\eta
_{E},\eta_{B}\rightarrow1$, and as $G\rightarrow1$. So $\mu_{\nu_{x}}%
-\alpha>-\theta$ if $-\theta+\alpha<\mu_{\nu_{x}}<\theta-\alpha$. Thus
$\sqrt{\eta_{E}\eta_{B}G}\left(  \mu_{\nu_{x}}-\alpha\right)  \geq\mu_{\nu
_{x}}-\alpha>-\theta$ whenever $\sqrt{\eta_{E}\eta_{B}G}<\theta/\left\vert
\mu_{\nu_{x}}-\alpha\right\vert $. Pick $\epsilon=\frac{1}{2}\left(
\theta-\sqrt{\eta_{E}\eta_{B}G}\alpha+\sqrt{\eta_{E}\eta_{B}G}\mu_{\nu_{x}%
}\right)  >0$. So $-\theta+\sqrt{\eta_{E}\eta_{B}G}\alpha+\epsilon=\sqrt
{\eta_{E}\eta_{B}G}\mu_{\nu_{x}}-\epsilon$. Then%
\begin{align*}
&  \int_{-\theta-\sqrt{\eta_{E}\eta_{B}G}\alpha}^{-\theta+\sqrt{\eta_{E}%
\eta_{B}G}\alpha}p_{N}\left(  n\right)  \ dn\\
&  \leq\int_{-\infty}^{-\theta+\sqrt{\eta_{E}\eta_{B}G}\alpha}p_{N}\left(
n\right)  \ dn\\
&  \leq\int_{-\infty}^{-\theta+\sqrt{\eta_{E}\eta_{B}G}\alpha+\epsilon}%
p_{N}\left(  n\right)  \ dn\\
&  \leq\int_{-\infty}^{\sqrt{\eta_{E}\eta_{B}G}\mu_{\nu_{x}}-\epsilon}%
p_{N}\left(  n\right)  \ dn\\
&  =\text{Pr}\left\{  N\leq\sqrt{\eta_{E}\eta_{B}G}\mu_{\nu_{x}}%
-\epsilon\right\}  \\
&  =\text{Pr}\left\{  N\leq\mu-\epsilon\right\}  \\
&  =\text{Pr}\left\{  N-\mu\leq-\epsilon\right\}  \\
&  \leq\text{Pr}\left\{  \left\vert N-\mu\right\vert \geq\epsilon\right\}  \\
&  \leq\frac{\sigma^{2}}{\epsilon^{2}}%
\end{align*}
So the conditional probability difference in (\ref{eq:cond-prob-diff-1})
vanishes as $\sigma_{\nu_{x}}^{2}\rightarrow0$, as $r\rightarrow\infty$, as
$\eta_{E},\eta_{B}\rightarrow1$, and as $G\rightarrow1$ when $-\theta
+\alpha<\mu_{\nu_{x}}<\theta-\alpha$. We now prove that the conditional
probability difference in (\ref{eq:cond-prob-diff-2}) vanishes as $\sigma
_{\nu_{x}}^{2}\rightarrow0$, as $r\rightarrow\infty$, as $\eta_{E},\eta
_{B}\rightarrow1$, and as $G\rightarrow1$. So $\mu_{\nu_{x}}+\alpha<\theta$ if
$-\theta+\alpha<\mu_{\nu_{x}}<\theta-\alpha$. Thus $\sqrt{\eta_{E}\eta_{B}%
G}\left(  \mu_{\nu_{x}}+\alpha\right)  \leq\mu_{\nu_{x}}+\alpha<\theta$
whenever $\sqrt{\eta_{E}\eta_{B}G}<\theta/\left\vert \mu_{\nu_{x}}%
+\alpha\right\vert $. Pick $\epsilon=\frac{1}{2}\left(  \theta-\sqrt{\eta
_{E}\eta_{B}G}\alpha-\sqrt{\eta_{E}\eta_{B}G}\mu_{\nu_{x}}\right)  >0$. So
$\theta-\sqrt{\eta_{E}\eta_{B}G}\alpha-\epsilon=\sqrt{\eta_{E}\eta_{B}G}%
\mu_{\nu_{x}}+\epsilon$.%
\begin{align*}
&  \int_{\theta-\sqrt{\eta_{E}\eta_{B}G}\alpha}^{\theta+\sqrt{\eta_{E}\eta
_{B}G}\alpha}p_{N}\left(  n\right)  \ dn\\
&  \leq\int_{\theta-\sqrt{\eta_{E}\eta_{B}G}\alpha}^{\infty}p_{N}\left(
n\right)  \ dn\\
&  \leq\int_{\theta-\sqrt{\eta_{E}\eta_{B}G}\alpha-\epsilon}^{\infty}%
p_{N}\left(  n\right)  \ dn\\
&  \leq\int_{\sqrt{\eta_{E}\eta_{B}G}\mu_{\nu_{x}}+\epsilon}^{\infty}%
p_{N}\left(  n\right)  \ dn\\
&  =\text{Pr}\left\{  N\geq\sqrt{\eta_{E}\eta_{B}G}\mu_{\nu_{x}}%
+\epsilon\right\}  \\
&  =\text{Pr}\left\{  N\geq\mu+\epsilon\right\}  \\
&  =\text{Pr}\left\{  N-\mu\geq\epsilon\right\}  \\
&  \leq\text{Pr}\left\{  \left\vert N-\mu\right\vert \geq\epsilon\right\}  \\
&  \leq\frac{\sigma^{2}}{\epsilon^{2}}%
\end{align*}
So the conditional probability difference in (\ref{eq:cond-prob-diff-2})
vanishes as $\sigma_{\nu_{x}}^{2}\rightarrow0$, as $r\rightarrow\infty$, as
$\eta_{E},\eta_{B}\rightarrow1$, and as $G\rightarrow1$ when $-\theta
+\alpha<\mu_{\nu_{x}}<\theta-\alpha$ and with the constraint $\sqrt{\eta
_{E}\eta_{B}G}\leq\min\left(  \theta/\left\vert \mu_{\nu_{x}}+\alpha
\right\vert ,\theta/\left\vert \mu_{\nu_{x}}-\alpha\right\vert \right)
$.\newline\newline Case 3: Suppose next that $\mu_{\nu_{x}}>\theta+\alpha$ so
that $\mu_{\nu_{x}}-\alpha>\theta>0$. We first prove that the conditional
probability difference in (\ref{eq:cond-prob-diff-1}) vanishes as $\sigma
_{\nu_{x}}^{2}\rightarrow0$, as $r\rightarrow\infty$, as $\eta_{E},\eta
_{B}\rightarrow1$, and as $G\rightarrow1$. So $\mu_{\nu_{x}}>-\theta+\alpha$
if $\mu_{\nu_{x}}>\theta+\alpha$. Thus $\mu_{\nu_{x}}>-\theta+\alpha$ and
$\mu_{\nu_{x}}-\alpha>-\theta$ and $\sqrt{\eta_{E}\eta_{B}G}\left(  \mu
_{\nu_{x}}-\alpha\right)  >-\theta$ for any $\sqrt{\eta_{E}\eta_{B}G}\geq0$.
Pick $\epsilon=\frac{1}{2}\left(  \sqrt{\eta_{E}\eta_{B}G}\mu_{\nu_{x}}%
+\theta-\sqrt{\eta_{E}\eta_{B}G}\alpha\right)  >0.$ So $-\theta+\sqrt{\eta
_{E}\eta_{B}G}\alpha+\epsilon=\sqrt{\eta_{E}\eta_{B}G}\mu_{\nu_{x}}-\epsilon.$
Then%
\begin{align*}
&  \int_{-\theta-\sqrt{\eta_{E}\eta_{B}G}\alpha}^{-\theta+\sqrt{\eta_{E}%
\eta_{B}G}\alpha}p_{N}\left(  n\right)  \ dn\\
&  \leq\int_{-\infty}^{-\theta+\sqrt{\eta_{E}\eta_{B}G}\alpha}p_{N}\left(
n\right)  \ dn\\
&  \leq\int_{-\infty}^{-\theta+\sqrt{\eta_{E}\eta_{B}G}\alpha+\epsilon}%
p_{N}\left(  n\right)  \ dn\\
&  \leq\int_{-\infty}^{\sqrt{\eta_{E}\eta_{B}G}\mu_{\nu_{x}}-\epsilon}%
p_{N}\left(  n\right)  \ dn\\
&  =\text{Pr}\left\{  N\leq\sqrt{\eta_{E}\eta_{B}G}\mu_{\nu_{x}}%
-\epsilon\right\}  \\
&  =\text{Pr}\left\{  N\leq\mu-\epsilon\right\}  \\
&  =\text{Pr}\left\{  N-\mu\leq-\epsilon\right\}  \\
&  \leq\text{Pr}\left\{  \left\vert N-\mu\right\vert \geq\epsilon\right\}  \\
&  \leq\frac{\sigma^{2}}{\epsilon^{2}}%
\end{align*}
So the conditional probability difference in (\ref{eq:cond-prob-diff-1})
vanishes as $\sigma_{\nu_{x}}^{2}\rightarrow0$, as $r\rightarrow\infty$, as
$\eta_{E},\eta_{B}\rightarrow1$, and as $G\rightarrow1$ when $\mu_{\nu_{x}%
}>\theta+\alpha$. We lastly prove that the conditional probability difference
in (\ref{eq:cond-prob-diff-2}) vanishes as $\sigma_{\nu_{x}}^{2}\rightarrow0$,
as $r\rightarrow\infty$, as $\eta_{E},\eta_{B}\rightarrow1$, and as
$G\rightarrow1$ when $\mu_{\nu_{x}}>\theta+\alpha$. So $\sqrt{\eta_{E}\eta
_{B}G}\left(  \mu_{\nu_{x}}-\alpha\right)  >\theta$ whenever $\sqrt{\eta
_{E}\eta_{B}G}>\theta/\left(  \mu_{\nu_{x}}-\alpha\right)  $. Pick
$\epsilon=\frac{1}{2}\left(  \sqrt{\eta_{E}\eta_{B}G}\mu_{\nu_{x}}%
-\theta-\sqrt{\eta_{E}\eta_{B}G}\alpha\right)  >0.$ So $\theta+\sqrt{\eta
_{E}\eta_{B}G}\alpha+\epsilon=\sqrt{\eta_{E}\eta_{B}G}\mu_{\nu_{x}}-\epsilon.$
Then%
\begin{align*}
&  \int_{\theta-\sqrt{\eta_{E}\eta_{B}G}\alpha}^{\theta+\sqrt{\eta_{E}\eta
_{B}G}\alpha}p_{N}\left(  n\right)  \ dn\\
&  \leq\int_{-\infty}^{\theta+\sqrt{\eta_{E}\eta_{B}G}\alpha}p_{N}\left(
n\right)  \ dn\\
&  \leq\int_{-\infty}^{\theta+\sqrt{\eta_{E}\eta_{B}G}\alpha+\epsilon}%
p_{N}\left(  n\right)  \ dn\\
&  \leq\int_{-\infty}^{\sqrt{\eta_{E}\eta_{B}G}\mu_{\nu_{x}}-\epsilon}%
p_{N}\left(  n\right)  \ dn\\
&  =\text{Pr}\left\{  N\leq\sqrt{\eta_{E}\eta_{B}G}\mu_{\nu_{x}}%
-\epsilon\right\}  \\
&  =\text{Pr}\left\{  N\leq\mu-\epsilon\right\}  \\
&  =\text{Pr}\left\{  N-\mu\leq-\epsilon\right\}  \\
&  \leq\text{Pr}\left\{  \left\vert N-\mu\right\vert \geq\epsilon\right\}  \\
&  \leq\frac{\sigma^{2}}{\epsilon^{2}}%
\end{align*}
So the conditional probability difference in (\ref{eq:cond-prob-diff-2}) as
$\sigma_{\nu_{x}}^{2}\rightarrow0$, as $r\rightarrow\infty$, and as $\eta
_{E},\eta_{B}\rightarrow1$ when $\mu_{\nu_{x}}>$ $\theta+\alpha$ and with the
constraint $\sqrt{\eta_{E}\eta_{B}G}>\theta/\left(  \mu_{\nu_{x}}%
-\alpha\right)  $. Thus $\mu_{\nu_{x}}\notin\left(  -\theta-\alpha
,-\theta+\alpha\right)  \cup\left(  \theta-\alpha,\theta+\alpha\right)  $ is a
sufficient condition for the nonmonotone SR effect to occur with the given
constraints on the product $\sqrt{\eta_{E}\eta_{B}G}$.
\end{proof}

\begin{proof}
[Proof (Necessity)]We prove that the SR\ effect does not occur when $\mu
_{\nu_{x}}\in\left(  -\theta-\alpha,-\theta+\alpha\right)  \cup\left(
\theta-\alpha,\theta+\alpha\right)  $.\newline\newline Case 1: Suppose first
that $\mu_{\nu_{x}}\in\left(  -\theta-\alpha,-\theta+\alpha\right)  $. We
prove with a similar Chebyshev bound that the conditional probabilities
$p_{Y|S}(0|0)\rightarrow1$ and $p_{Y|S}(\varepsilon|1)\rightarrow1$ as
$\sigma_{\nu_{x}}^{2}\rightarrow0$, as $r\rightarrow\infty$, as $\eta_{E}%
,\eta_{B}\rightarrow1$, and as $G\rightarrow1$. Then the mutual information
$I(S,Y)$ approaches its maximum $H\left(  S\right)  $ as all noise vanishes.
Consider $p_{Y|S}(0|0)$. Pick any $\mu_{\nu_{x}}\in\left(  -\theta
-\alpha,-\theta+\alpha\right)  $. Then $-\theta+\alpha>\mu_{\nu_{x}}$ and
$\alpha-\mu_{\nu_{x}}>\theta$. Then $-\theta>\sqrt{\eta_{E}\eta_{B}G}\left(
\mu_{\nu_{x}}-\alpha\right)  $ whenever $\sqrt{\eta_{E}\eta_{B}G}%
>\theta/\left\vert \alpha-\mu_{\nu_{x}}\right\vert $. Pick $\epsilon=\frac
{1}{2}\left(  -\theta+\sqrt{\eta_{E}\eta_{B}G}\alpha-\sqrt{\eta_{E}\eta_{B}%
G}\mu_{\nu_{x}}\right)  >0$ so that $-\theta+\sqrt{\eta_{E}\eta_{B}G}%
\alpha-\epsilon=\sqrt{\eta_{E}\eta_{B}G}\mu_{\nu_{x}}+\epsilon$.%
\begin{align*}
&  p_{Y|S}(0|0)\\
&  =\int_{-\infty}^{-\theta+\sqrt{\eta_{E}\eta_{B}G}\alpha}p_{N}\left(
n\right)  \ dn\\
&  \geq\int_{-\infty}^{-\theta+\sqrt{\eta_{E}\eta_{B}G}\alpha-\epsilon}%
p_{N}\left(  n\right)  \ dn\\
&  =\int_{-\infty}^{\sqrt{\eta_{E}\eta_{B}G}\mu_{\nu_{x}}+\epsilon}%
p_{N}\left(  n\right)  \ dn\\
&  =1-\int_{\sqrt{\eta_{E}\eta_{B}G}\mu_{\nu_{x}}+\epsilon}^{\infty}%
p_{N}\left(  n\right)  \ dn\\
&  =1-\text{Pr}\left\{  N\geq\sqrt{\eta_{E}\eta_{B}G}\mu_{\nu_{x}}%
+\epsilon\right\}  \\
&  =1-\text{Pr}\left\{  N\geq\mu+\epsilon\right\}  \\
&  =1-\text{Pr}\left\{  N-\mu\geq\epsilon\right\}  \\
&  \geq1-\text{Pr}\left\{  \left\vert N-\mu\right\vert \geq\epsilon\right\}
\\
&  \geq1-\frac{\sigma^{2}}{\epsilon^{2}}\\
&  \rightarrow1
\end{align*}
as $\sigma_{\nu_{x}}^{2}\rightarrow0$, as $r\rightarrow\infty$, as $\eta
_{E},\eta_{B}\rightarrow1$, and as $G\rightarrow1$. We prove the result
similarly for $p_{Y|S}(\varepsilon|1)$. We show that $p_{Y|S}(0|1)\rightarrow
0$ and $p_{Y|S}(1|1)\rightarrow0$ so that $p_{Y|S}(\varepsilon|1)\rightarrow
1$. Pick any $\mu_{\nu_{x}}\in\left(  -\theta-\alpha,-\theta+\alpha\right)  $.
Then $\mu_{\nu_{x}}>-\theta-\alpha$ and $\mu_{\nu_{x}}+\alpha>-\theta$.
$\sqrt{\eta_{E}\eta_{B}G}\left(  \mu_{\nu_{x}}+\alpha\right)  >-\theta$ and
$\sqrt{\eta_{E}\eta_{B}G}\mu_{\nu_{x}}+\sqrt{\eta_{E}\eta_{B}G}\alpha
+\theta>0$ whenever $\sqrt{\eta_{E}\eta_{B}G}<\theta/\left\vert \mu_{\nu_{x}%
}+\alpha\right\vert $. Pick $\epsilon=\frac{1}{2}\left(  \sqrt{\eta_{E}%
\eta_{B}G}\mu_{\nu_{x}}+\sqrt{\eta_{E}\eta_{B}G}\alpha+\theta\right)  >0$ so
that $-\theta-\sqrt{\eta_{E}\eta_{B}G}\alpha+\epsilon=\sqrt{\eta_{E}\eta_{B}%
G}\mu_{\nu_{x}}-\epsilon$.%
\begin{align*}
&  p_{Y|S}(0|1)\\
&  =\int_{-\infty}^{-\theta-\sqrt{\eta_{E}\eta_{B}G}\alpha}p_{N}\left(
n\right)  \ dn\\
&  \leq\int_{-\infty}^{-\theta-\sqrt{\eta_{E}\eta_{B}G}\alpha+\epsilon}%
p_{N}\left(  n\right)  \ dn\\
&  =\int_{-\infty}^{\sqrt{\eta_{E}\eta_{B}G}\mu_{\nu_{x}}-\epsilon}%
p_{N}\left(  n\right)  \ dn\\
&  =\text{Pr}\left\{  N\leq\sqrt{\eta_{E}\eta_{B}G}\mu_{\nu_{x}}%
-\epsilon\right\}  \\
&  =\text{Pr}\left\{  N\leq\mu-\epsilon\right\}  \\
&  =\text{Pr}\left\{  N-\mu\leq-\epsilon\right\}  \\
&  \leq\text{Pr}\left\{  \left\vert N-\mu\right\vert \geq\epsilon\right\}  \\
&  \leq\frac{\sigma^{2}}{\epsilon^{2}}%
\end{align*}
Pick any $\mu_{\nu_{x}}\in\left(  -\theta-\alpha,-\theta+\alpha\right)  $.
Then $\mu_{\nu_{x}}<-\theta+\alpha$ and $\theta<\alpha-\mu_{\nu_{x}}$.
$\sqrt{\eta G}\left(  \mu_{\nu_{x}}-\alpha\right)  <-\theta$ and $-\sqrt
{\eta_{E}\eta_{B}G}\mu_{\nu_{x}}+\sqrt{\eta_{E}\eta_{B}G}\alpha-\theta>0$
whenever $\sqrt{\eta_{E}\eta_{B}G}>\theta/\left\vert \alpha-\mu_{\nu_{x}%
}\right\vert $. Pick $\epsilon=\frac{1}{2}\left(  -\sqrt{\eta_{E}\eta_{B}G}%
\mu_{\nu_{x}}+\sqrt{\eta_{E}\eta_{B}G}\alpha-\theta\right)  >0$ so that
$-\theta+\sqrt{\eta_{E}\eta_{B}G}\alpha-\epsilon=\sqrt{\eta_{E}\eta_{B}G}%
\mu_{\nu_{x}}+\epsilon$.%
\begin{align*}
&  p_{Y|S}(1|1)\\
&  =\int_{\theta-\sqrt{\eta_{E}\eta_{B}G}\alpha}^{\infty}p_{N}\left(
n\right)  \ dn\\
&  \leq\int_{-\theta+\sqrt{\eta_{E}\eta_{B}G}\alpha}^{\infty}p_{N}\left(
n\right)  \ dn\\
&  \leq\int_{-\theta+\sqrt{\eta_{E}\eta_{B}G}\alpha-\epsilon}^{\infty}%
p_{N}\left(  n\right)  \ dn\\
&  =\int_{\sqrt{\eta_{E}\eta_{B}G}\mu_{\nu_{x}}+\epsilon}^{\infty}p_{N}\left(
n\right)  \ dn\\
&  =\text{Pr}\left\{  N\geq\sqrt{\eta_{E}\eta_{B}G}\mu_{\nu_{x}}%
+\epsilon\right\}  \\
&  =\text{Pr}\left\{  N\geq\mu+\epsilon\right\}  \\
&  =\text{Pr}\left\{  N-\mu\geq\epsilon\right\}  \\
&  \leq\text{Pr}\left\{  \left\vert N-\mu\right\vert \geq\epsilon\right\}  \\
&  \leq\frac{\sigma^{2}}{\epsilon^{2}}%
\end{align*}
So $p_{Y|S}(\varepsilon|1)\rightarrow1$ because $p_{Y|S}(0|1)\rightarrow0$ and
$p_{Y|S}(1|1)\rightarrow0$ as $\sigma_{\nu_{x}}^{2}\rightarrow0$, as
$r\rightarrow\infty$, as $\eta_{E},\eta_{B}\rightarrow1$, and as
$G\rightarrow1$ and with constraint $\theta/\left\vert \alpha-\mu_{\nu_{x}%
}\right\vert <\sqrt{\eta_{E}\eta_{B}G}<\theta/\left\vert \alpha+\mu_{\nu_{x}%
}\right\vert $.\newline\newline Case 2: Now suppose that $\mu_{\nu_{x}}%
\in\left(  \theta-\alpha,\theta+\alpha\right)  $. We prove that the
conditional probabilities $p_{Y|S}(\varepsilon|0)\rightarrow1$ and
$p_{Y|S}(1|1)\rightarrow1$ as $\sigma_{\nu_{x}}^{2}\rightarrow0$, as
$r\rightarrow\infty$, as $\eta_{E},\eta_{B}\rightarrow1$, and as
$G\rightarrow1$. We first prove that $p_{Y|S}(\varepsilon|0)\rightarrow1$ in
the limit of zero noise. We prove this by showing that $p_{Y|S}%
(0|0)\rightarrow0$ and $p_{Y|S}(1|0)\rightarrow0$ in the limit. Pick any
$\mu_{\nu_{x}}\in\left(  \theta-\alpha,\theta+\alpha\right)  $. Then $\mu
_{\nu_{x}}>\theta-\alpha$ and $\mu_{\nu_{x}}+\alpha>\theta$. $\sqrt{\eta
_{E}\eta_{B}G}\left(  \mu_{\nu_{x}}+\alpha\right)  >\theta$ and $\sqrt
{\eta_{E}\eta_{B}G}\mu_{\nu_{x}}+\sqrt{\eta_{E}\eta_{B}G}\alpha-\theta>0$
whenever $\sqrt{\eta_{E}\eta_{B}G}>\theta/\left(  \mu_{\nu_{x}}+\alpha\right)
$. Pick $\epsilon=\frac{1}{2}\left(  \sqrt{\eta_{E}\eta_{B}G}\mu_{\nu_{x}%
}+\sqrt{\eta_{E}\eta_{B}G}\alpha-\theta\right)  >0$ so that $\theta-\sqrt
{\eta_{E}\eta_{B}G}\alpha+\epsilon=\sqrt{\eta_{E}\eta_{B}G}\mu_{\nu_{x}%
}-\epsilon$.%
\begin{align*}
&  p_{Y|S}(0|0)\\
&  =\int_{-\infty}^{-\theta+\sqrt{\eta_{E}\eta_{B}G}\alpha}p_{N}\left(
n\right)  \ dn\\
&  \leq\int_{-\infty}^{\theta-\sqrt{\eta_{E}\eta_{B}G}\alpha}p_{N}\left(
n\right)  \ dn\\
&  \leq\int_{-\infty}^{\theta-\sqrt{\eta_{E}\eta_{B}G}\alpha+\epsilon}%
p_{N}\left(  n\right)  \ dn\\
&  =\int_{-\infty}^{\sqrt{\eta_{E}\eta_{B}G}\mu_{\nu_{x}}-\epsilon}%
p_{N}\left(  n\right)  \ dn\\
&  =\text{Pr}\left\{  N\leq\sqrt{\eta_{E}\eta_{B}G}\mu_{\nu_{x}}%
-\epsilon\right\}  \\
&  =\text{Pr}\left\{  N\leq\mu-\epsilon\right\}  \\
&  =\text{Pr}\left\{  N-\mu\leq-\epsilon\right\}  \\
&  \leq\text{Pr}\left\{  \left\vert N-\mu\right\vert \geq\epsilon\right\}  \\
&  \leq\frac{\sigma^{2}}{\epsilon^{2}}%
\end{align*}
Pick any $\mu_{\nu_{x}}\in\left(  \theta-\alpha,\theta+\alpha\right)  $. Then
$\mu_{\nu_{x}}<\theta+\alpha$ and $\mu_{\nu_{x}}-\alpha<\theta$. $\sqrt
{\eta_{E}\eta_{B}G}\left(  \mu_{\nu_{x}}-\alpha\right)  <\theta$ and
$-\sqrt{\eta_{E}\eta_{B}G}\mu_{\nu_{x}}+\sqrt{\eta_{E}\eta_{B}G}\alpha
+\theta>0$ whenever $\sqrt{\eta_{E}\eta_{B}G}<\theta/\left\vert \mu_{\nu_{x}%
}-\alpha\right\vert $. Pick $\epsilon=\frac{1}{2}\left(  -\sqrt{\eta_{E}%
\eta_{B}G}\mu_{\nu_{x}}+\sqrt{\eta_{E}\eta_{B}G}\alpha+\theta\right)  >0$ so
that $\theta+\sqrt{\eta_{E}\eta_{B}G}\alpha-\epsilon=\sqrt{\eta_{E}\eta_{B}%
G}\mu_{\nu_{x}}+\epsilon$.%
\begin{align*}
&  p_{Y|S}(1|0)\\
&  =\int_{\theta+\sqrt{\eta_{E}\eta_{B}G}\alpha}^{\infty}p_{N}\left(
n\right)  \ dn\\
&  \leq\int_{\theta+\sqrt{\eta_{E}\eta_{B}G}\alpha-\epsilon}^{\infty}%
p_{N}\left(  n\right)  \ dn\\
&  =\int_{\sqrt{\eta_{E}\eta_{B}G}\mu_{\nu_{x}}+\epsilon}^{\infty}p_{N}\left(
n\right)  \ dn\\
&  =\text{Pr}\left\{  N\geq\sqrt{\eta_{E}\eta_{B}G}\mu_{\nu_{x}}%
+\epsilon\right\}  \\
&  =\text{Pr}\left\{  N\geq\mu+\epsilon\right\}  \\
&  =\text{Pr}\left\{  N-\mu\geq\epsilon\right\}  \\
&  \leq\text{Pr}\left\{  \left\vert N-\mu\right\vert \geq\epsilon\right\}  \\
&  \leq\frac{\sigma^{2}}{\epsilon^{2}}%
\end{align*}
So $p_{Y|S}(\varepsilon|0)\rightarrow1$ because $p_{Y|S}(0|0)\rightarrow0$ and
$p_{Y|S}(1|0)\rightarrow0$ as $\sigma_{\nu_{x}}^{2}\rightarrow0$, as
$r\rightarrow\infty$, as $\eta_{E},\eta_{B}\rightarrow1$, and as
$G\rightarrow1$. Now we prove that $p_{Y|S}(1|1)\rightarrow1$ as $\sigma
_{\nu_{x}}^{2}\rightarrow0$, as $r\rightarrow\infty$, as $\eta_{E},\eta
_{B}\rightarrow1$, and as $G\rightarrow1$ whenever $\mu_{\nu_{x}}\in\left(
\theta-\alpha,\theta+\alpha\right)  $. Pick any $\mu_{\nu_{x}}\in\left(
\theta-\alpha,\theta+\alpha\right)  $. Then $\mu_{\nu_{x}}<\theta+\alpha$ and
$\mu_{\nu_{x}}-\alpha<\theta$. $\sqrt{\eta_{E}\eta_{B}G}\left(  \mu_{\nu_{x}%
}-\alpha\right)  <\theta$ and $-\sqrt{\eta_{E}\eta_{B}G}\mu_{\nu_{x}}%
+\sqrt{\eta_{E}\eta_{B}G}\alpha+\theta>0$ whenever $\sqrt{\eta_{E}\eta_{B}%
G}<\theta/\left\vert \mu_{\nu_{x}}-\alpha\right\vert $. Pick $\epsilon
=\frac{1}{2}\left(  -\sqrt{\eta_{E}\eta_{B}G}\mu_{\nu_{x}}+\sqrt{\eta_{E}%
\eta_{B}G}\alpha+\theta\right)  >0$ so that $\theta-\sqrt{\eta_{E}\eta_{B}%
G}\alpha+\epsilon=\sqrt{\eta_{E}\eta_{B}G}\mu_{\nu_{x}}-\epsilon$.%
\begin{align*}
&  p_{Y|S}(1|1)\\
&  =\int_{\theta-\sqrt{\eta_{E}\eta_{B}G}\alpha}^{\infty}p_{N}\left(
n\right)  \ dn\\
&  \geq\int_{\theta-\sqrt{\eta_{E}\eta_{B}G}\alpha+\epsilon}^{\infty}%
p_{N}\left(  n\right)  \ dn\\
&  =\int_{\sqrt{\eta_{E}\eta_{B}G}\mu_{\nu_{x}}-\epsilon}^{\infty}p_{N}\left(
n\right)  \ dn\\
&  =1-\int_{-\infty}^{\sqrt{\eta_{E}\eta_{B}G}\mu_{\nu_{x}}-\epsilon}%
p_{N}\left(  n\right)  \ dn\\
&  =1-\text{Pr}\left\{  N\leq\sqrt{\eta_{E}\eta_{B}G}\mu_{\nu_{x}}%
-\epsilon\right\}  \\
&  =1-\text{Pr}\left\{  N\leq\mu-\epsilon\right\}  \\
&  =1-\text{Pr}\left\{  N-\mu\leq-\epsilon\right\}  \\
&  \geq1-\text{Pr}\left\{  \left\vert N-\mu\right\vert \geq\epsilon\right\}
\\
&  \geq1-\frac{\sigma^{2}}{\epsilon^{2}}%
\end{align*}
So $p_{Y|S}(1|1)\rightarrow1$ as $\sigma_{\nu_{x}}^{2}\rightarrow0$, as
$r\rightarrow\infty$, as $\eta_{E},\eta_{B}\rightarrow1$, and as
$G\rightarrow1$ whenever $\mu_{\nu_{x}}\in\left(  \theta-\alpha,\theta
+\alpha\right)  $ and with constraint $\theta/\left(  \mu_{\nu_{x}}%
+\alpha\right)  <\sqrt{\eta_{E}\eta_{B}G}<\theta/\left\vert \mu_{\nu_{x}%
}-\alpha\right\vert $. The mutual information $I(S,Y)$ approaches its maximum
$H\left(  S\right)  $ as all noise vanishes and the SR\ effect does not occur
for Alice and Bob's mutual information whenever $\mu_{\nu_{x}}\in\left(
-\theta-\alpha,-\theta+\alpha\right)  \cup\left(  \theta-\alpha,\theta
+\alpha\right)  $ with the above constraints on the product $\sqrt{\eta
_{E}\eta_{B}G}$.\pagebreak
\end{proof}

\subsection{Proof of Theorem \ref{thm:qkd-finite-var-QFIT} (Infinite
Variance)}

\label{sec:proof-CVQKD-QFIT-infinite-variance}The proof for sufficiency and
necessity follows the same stable proof method with some modifications. We use
the same characteristic function $\varphi_{\nu_{x}}\left(  \omega\right)  $ in
(\ref{eq:characteristic-alpha-stable}) for alpha-stable random variable
$\nu_{x}$. Suppose%
\begin{equation}
\sigma_{\mathcal{N}}^{2}=\left(  \eta_{B}\left(
\begin{array}
[c]{c}%
\eta_{E}Ge^{-2r}+\eta_{E}\left(  G-1\right)  \\
+\left(  1-\eta_{E}\right)
\end{array}
\right)  +1-\eta_{B}\right)  /2.
\end{equation}
The characteristic function $\varphi_{N}\left(  \omega\right)  $ of
$p_{N}\left(  n\right)  $ is as follows%
\begin{equation}
\varphi_{N}\left(  \omega\right)  =\varphi_{\nu_{x}}\left(  \sqrt{\eta_{E}%
\eta_{B}G}\omega\right)  \exp\left\{  -\frac{\sigma_{\mathcal{N}}^{2}%
\omega^{2}}{2}\right\}
\end{equation}
from (\ref{eq:noise-density}) and the convolution theorem.

\begin{proof}
[Proof (Sufficiency)]Take the limit of the characteristic function
$\varphi_{N}\left(  \omega\right)  $ as $\gamma\rightarrow0$, as
$r\rightarrow\infty$, as $G\rightarrow1$, and as $\eta_{E},\eta_{B}%
\rightarrow1$ to obtain the following characteristic function.%
\begin{equation}
\lim_{\substack{r\rightarrow\infty,\gamma\rightarrow0,\\G\rightarrow1,\eta
_{E},\eta_{B}\rightarrow1}}\varphi_{N}(\omega)=\exp\left\{  ia\omega\right\}
\end{equation}
The probability density $p_{N}\left(  n\right)  $ then approaches a translated
delta function%
\begin{equation}
\lim_{r\rightarrow\infty,\gamma\rightarrow0,\eta_{E},\eta_{B}\rightarrow
1}p_{N}\left(  n\right)  =\delta\left(  n-a\right)
\end{equation}
Suppose that $a\notin\left(  -\theta-\alpha,-\theta+\alpha\right)  \cup\left(
\theta-\alpha,\theta+\alpha\right)  $. Consider the case where $y=0$. Then%
\begin{align}
&  \lim_{\substack{r\rightarrow\infty,\gamma\rightarrow0,\\G\rightarrow
1,\eta_{E},\eta_{B}\rightarrow1}}p_{Y|S}(0|0)-p_{Y|S}(0|1)\\
&  =\lim_{\substack{r\rightarrow\infty,\gamma\rightarrow0,\\G\rightarrow
1,\eta_{E},\eta_{B}\rightarrow1}}\int_{-\theta-\sqrt{\eta_{E}\eta_{B}G}\alpha
}^{-\theta+\sqrt{\eta_{E}\eta_{B}G}\alpha}p_{N}\left(  n\right)  \ dn\\
&  \rightarrow\int_{-\theta-\alpha}^{-\theta+\alpha}\delta\left(  n-a\right)
\ dn=0
\end{align}
because $a\notin\left(  -\theta-\alpha,-\theta+\alpha\right)  $. Consider the
case where $y=1$.%
\begin{align}
&  \lim_{\substack{r\rightarrow\infty,\gamma\rightarrow0,\\G\rightarrow
1,\eta_{E},\eta_{B}\rightarrow1}}p_{Y|S}(1|1)-p_{Y|S}(1|0)\\
&  =\lim_{\substack{r\rightarrow\infty,\gamma\rightarrow0,\\G\rightarrow
1,\eta_{E},\eta_{B}\rightarrow1}}\int_{\theta-\sqrt{\eta_{E}\eta_{B}G}\alpha
}^{\theta+\sqrt{\eta_{E}\eta_{B}G}\alpha}p_{N}\left(  n\right)  \ dn\\
&  \rightarrow\int_{\theta-\alpha}^{\theta+\alpha}\delta\left(  n-a\right)
\ dn=0
\end{align}
because $a\notin\left(  \theta-\alpha,\theta+\alpha\right)  $.
\end{proof}

\begin{proof}
[Proof (Necessity)]Suppose that $a\in\left(  -\theta-\alpha,-\theta
+\alpha\right)  \cup\left(  \theta-\alpha,\theta+\alpha\right)  $%
.\newline\newline Case 1:\ Pick $a\in\left(  -\theta-\alpha,-\theta
+\alpha\right)  $. We show that $p_{Y|S}(0|0)\rightarrow1$ and $p_{Y|S}%
(\varepsilon|1)\rightarrow1$. Then%
\begin{align}
p_{Y|S}(0|0)  &  =\int_{-\infty}^{-\theta+\sqrt{\eta_{E}\eta_{B}G}\alpha_{x}%
}p_{N}\left(  n\right)  \ dn\\
&  =\int_{-\infty}^{-\theta}p_{N}\left(  n+\sqrt{\eta_{E}\eta_{B}G}\alpha
_{x}\right)  \ dn\\
&  \rightarrow\int_{-\infty}^{-\theta}\delta\left(  n-a+\alpha\right)  \ dn\\
&  =\int_{-\infty}^{-\theta+\alpha}\delta\left(  n-a\right)  \ dn=1
\end{align}%
\begin{align}
p_{Y|S}(\varepsilon|1)  &  =\int_{-\theta-\sqrt{\eta_{E}\eta_{B}G}\alpha
}^{\theta-\sqrt{\eta_{E}\eta_{B}G}\alpha}p_{N}\left(  n\right)  \ dn\\
&  \rightarrow\int_{-\theta-\alpha}^{\theta-\alpha}\delta\left(  n-a\right)
\ dn\\
&  =1
\end{align}
Case 2:\ Pick $a\in\left(  \theta-\alpha,\theta+\alpha\right)  $. We show that
$p_{Y|S}(1|1)\rightarrow1$ and $p_{Y|S}(\varepsilon|0)\rightarrow1$. Then%
\begin{align}
p_{Y|S}(1|1)  &  =\int_{\theta-\sqrt{\eta_{E}\eta_{B}G}\alpha}^{\infty}%
p_{N}\left(  n\right)  \ dn\\
&  =\int_{\theta}^{\infty}p_{N}\left(  n-\sqrt{\eta_{E}\eta_{B}G}\alpha
_{x}\right)  \ dn\\
&  \rightarrow\int_{\theta}^{\infty}\delta\left(  n-a-\alpha\right)  \ dn\\
&  =\int_{\theta-\alpha}^{\infty}\delta\left(  n-a\right)  \ dn=1
\end{align}%
\begin{align}
p_{Y|S}(\varepsilon|0)  &  =\int_{-\theta+\sqrt{\eta_{E}\eta_{B}G}\alpha
}^{\theta+\sqrt{\eta_{E}\eta_{B}G}\alpha}p_{N}\left(  n\right)  \ dn\\
&  \rightarrow\int_{-\theta+\alpha}^{\theta+\alpha}\delta\left(  n-a\right)
\ dn\\
&  =1
\end{align}

\end{proof}

\bibliographystyle{apsrev}
\bibliography{qsr-prl}

\end{document}